\documentclass[a4paper,fleqn,usenatbib]{mnras}

\usepackage[T1]{fontenc}
\usepackage{ae,aecompl}

\usepackage{color}
\usepackage{graphicx}	
\usepackage{amsmath}	
\usepackage{amssymb}	

\title[Young TDGs in nearby groups]{The frequency and properties of young tidal dwarf galaxies in nearby gas-rich groups}

\author[K. Lee-Waddell et al.]
	{K.~Lee-Waddell$^{1}$\thanks{E-mail: karen.lee-waddell@csiro.au}, 
	K.~Spekkens$^{2}$, P.~Chandra$^{3}$, N.~Patra$^{3}$, J.-C.~Cuillandre$^{4}$, 
\newauthor 
	J.~Wang$^{1}$, M.~P.~Haynes$^{5}$, J.~Cannon$^{6}$, S.~Stierwalt$^{7}$, J.~Sick$^{8}$ and R.~Giovanelli$^{5}$
\\
$^{1}$CSIRO Astronomy and Space Sciences, Australia Telescope National Facility, PO Box 76, Epping, NSW 1710, Australia\\	
$^{2}$Department of Physics, Royal Military College of Canada, PO Box 17000, Station Forces, Kingston, ON K7K 7B4, Canada\\
$^{3}$National Centre for Radio Astrophysics, Tata Institute of Fundamental Research, Pune 411 007, India\\
$^{4}$CEA/IRFU/SAp, Laboratoire AIM Paris-Saclay, CNRS/INSU, Universit\'{e} Paris Diderot, Observatoire de Paris, PSL Research \\
\hspace{1.3mm}University, F-91191 Gif-sur-Yvette Cedex, France\\
$^{5}$Cornell Center for Astrophysics and Planetary Science, Cornell University, Ithaca, NY 14853, USA\\
$^{6}$Department of Physics and Astronomy, Macalester College, 1600 Grand Avenue, Saint Paul, MN 55105, USA\\
$^{7}$Department of Astronomy, University of Virginia, 530 McCormick Road, Charlottesville, VA 22904, USA\\
$^{8}$Department of Physics, Engineering Physics and Astronomy, Queen's University, Kingston, ON K7L 3N6, Canada
}

\date{Accepted 2016 May 12. Received 2016 May 10; in original form 2015 November 04}

\pubyear{2016}

\begin{document}
\label{firstpage}
\pagerange{\pageref{firstpage}--\pageref{lastpage}}
\maketitle

\begin{abstract}
We present high-resolution Giant Metrewave Radio Telescope (GMRT) {H\sc{i}} observations and deep Canada-France-Hawaii Telescope (CFHT) optical imaging of two galaxy groups: NGC 4725/47 and NGC 3166/9.  These data are part of a multi-wavelength unbiased survey of the gas-rich dwarf galaxy populations in three nearby interacting galaxy groups.  The NGC~4725/47 group hosts two tidal knots and one dIrr.  Both tidal knots are located within a prominent {H\sc{i}} tidal tail, appear to have sufficient mass ($M_{gas}$ $\approx$ 10$^8$ $M_{\odot}$) to evolve into long-lived tidal dwarf galaxies (TDGs) and are fairly young in age.  The NGC~3166/9 group contains a TDG candidate, AGC~208457, at least three dIrrs and four {H\sc{i}} knots.  Deep CFHT imaging confirms that the optical component of AGC~208457 is bluer --- with a 0.28 mag $g$-$r$ colour --- and a few Gyr younger than its purported parent galaxies.  Combining the results for these groups with those from the NGC~871/6/7 group reported earlier, we find that the {H\sc{i}} properties, estimated stellar ages and baryonic content of the gas-rich dwarfs clearly distinguish tidal features from their classical counterparts.  We optimistically identify four potentially long-lived tidal objects associated to three separate pairs of interacting galaxies, implying that TDGs are not readily produced during interaction events as suggested by some recent simulations.  The tidal objects examined in this survey also appear to have a wider variety of properties than TDGs of similar mass formed in current simulations of interacting galaxies, which could be the result of pre- or post-formation environmental influences.
\end{abstract}




\section{Introduction}
\label{intro}
Most present-day galaxies reside in medium-density group environments (\citealt{eke2004}, \citealt{tag2008}), where tidal interactions dominate the dynamics of the contained members.  Gaseous material pulled from initial close encounters between galaxies can produce second-generation tidal dwarf galaxies (TDGs), which differ from first-generation `classical' dwarfs by their lack of dark matter, increased star formation rates (SFRs) and higher metallicity due to pre-enriched stars drawn from parent galaxies  (\citealt{hun2000}, \citealt{duc2007}, \citealt{kav2012}, \citealt{lel2015}).  The overall prevalence of TDGs and their significance within the dwarf galaxy population are hotly debated topics (e.g.~\citealt{kro2012}, \citealt{duc2014}).  

Hierarchical galaxy formation simulations imply that the incidence of TDGs should be quite low at the current epoch \citep{sh2009}; whereas, numerical modelling and extrapolation by \citet{dab2013} conclude that the majority of nearby galaxies should be tidal in nature.  Sample size is one of the main sources of discrepancy between various studies.  For example, \citet{kro2012} uses three TDGs to argue that the similarities in the properties between tidal and classical dwarfs disproves the existence of dark matter.  Additionally, very few ($<20$) TDGs are widely considered as authentic \citep{w2003} due to the variety of corroborating observations that are required to confirm that classification \citep{duc2014}.  

The formation of tidal knots --- i.e.~clumps of gas and possibly stars located near or within tidal tails that do not show signs of self-gravitation and are therefore not (yet) considered to be actual galaxies --- also appears to skew each argument.  Numerical simulations of a multitude of paired spiral galaxy interactions by \citet{bou2006} show that $\sim$6.2 tidal knots (with $M_{baryon} \geq 10^8$ $M_{\odot}$) are formed per interaction event.  The majority of these knots fall back into their parents within $\sim$10$^2$ Myr and by the end of the simulations (t = 2 Gyr) $\sim$1.2 long-lived tidal knots, which at that point are considered TDGs, remain.  Conversely, simulations used by \citet{yan2014} and references therein --- in an attempt to explain the planes of satellite galaxies detected in the Local Group (LG; see \citealt{paw2012}, \citealt{iba2013}) --- suggest that a few tens of tidal knots can be formed in each tidal filament produced during an encounter, and that each knot has adequate mass ($M_{baryon} \geq 10^8$ $M_{\odot}$) and separation distance from the gravitational well of their parent galaxies to survive for several Gyr.

We have conducted a detailed investigation of three nearby gas-rich interacting galaxy groups to study the properties and frequency of TDGs in these environments, affording an opportunity to place observational constraints on the different TDG formation and survival rates obtained in the aforementioned simulations.  The groups were selected from the single-dish neutral hydrogen ({H\sc{i}}) Arecibo Legacy Fast ALFA survey (ALFALFA; \citealt{gio2005}) and show signs of recent or on-going tidal interaction events dominated by a pair of spiral galaxies, which optimizes likelihood of detecting young TDGs.  The blind nature of ALFALFA allowed us to follow up every gas-rich detection in these groups in order to determine its origin, thereby producing an unbiased census of first- and second-generation systems. 

We have obtained high-resolution follow-up observations for each group with the Giant Metrewave Radio Telescope (GMRT) to identify gas-rich dwarfs and measure their {H\sc{i}} and dynamical masses (see \citealt{lee2012}; hereafter Paper~1).  Deep optical imaging, from the Canada-France-Hawaii Telescope (CFHT) MegaCam, has been used to identify putative low surface brightness optical counterparts of the low-mass features and estimate their stellar masses and ages (see \citealt{lee2014}; hereafter Paper~2).  The combination of our {H\sc{i}} imaging and optical photometry allows for dynamical to baryonic mass calculations and stellar population estimates that we use to classify TDG candidates, short-lived tidal knots and first-generation gas-rich dwarf irregular galaxies (dIrrs) in our census of gas-rich dwarf galaxies in three nearby groups.  

NGC~3166/9 is a gas-rich group that comprises three spiral galaxies and shows signs of recent tidal interactions.  In Paper~1, we detected and resolved the atomic gas component of eight low-mass group members including a TDG candidate, AGC~208457, using a mosaic of six GMRT pointings.  That paper outlines our methodology for analyzing interferometric {H\sc{i}} observations.  The GMRT data for the NGC~3166/9 group indicate that AGC~208457 has sufficient mass to be self-gravitating and has a dynamical to gas mass ratio close to unity, which denotes a lack of dark matter and is one of the hallmarks of a TDG.  Given the {H\sc{i}} and SDSS-depth optical properties of the remaining seven detections, we classified three as typical dIrrs and the remaining four as short-lived tidal knots that will likely fall back into their parent galaxies.

NGC~871/6/7 also consists of three gas-rich spirals that are located within an extended {H\sc{i}} distribution, which indicates ongoing tidal interactions within this group.  In Paper~2, our {H\sc{i}} observations revealed five gas-rich low-mass objects, four of which are likely dIrrs.  The other low-mass feature is AGC~749170, which has an {H\sc{i}} mass $M_{H_I} \sim 1.4\times 10^{9}$ $M_{\odot}$ but an extremely faint stellar counterpart near the detection limit of our deep CFHT data, implying a dynamical mass to stellar light ratio $M_{dyn}/L_{g}>1000$ $M_{\odot}/L_{\odot}$.  The lack of dark matter ($M_{dyn}/M_{H_I} \sim 1$) associated with AGC~749170 and the apparently young age of the putative optical counterpart suggest tidal origins for this intriguing gas cloud; however, while it is massive enough to be long-lived (e.g.~\citealt{bou2006}), its low stellar content does not resemble any of the tidal systems obtained in simulations.  Note, the extremely optically dim optical feature that spatially coincides with AGC~749170 was detected, for the first time, by MegaCam and the Elixir/Elixir-LSB processing pipeline (\citealt{mag2004}, \citealt{cui2011}), which illustrates the utility of deep optical photometry for understanding the origin of {H\sc{i}} features in nearby interacting galaxy groups. 

\begin{figure}
\begin{center}
  \includegraphics[width=84mm]{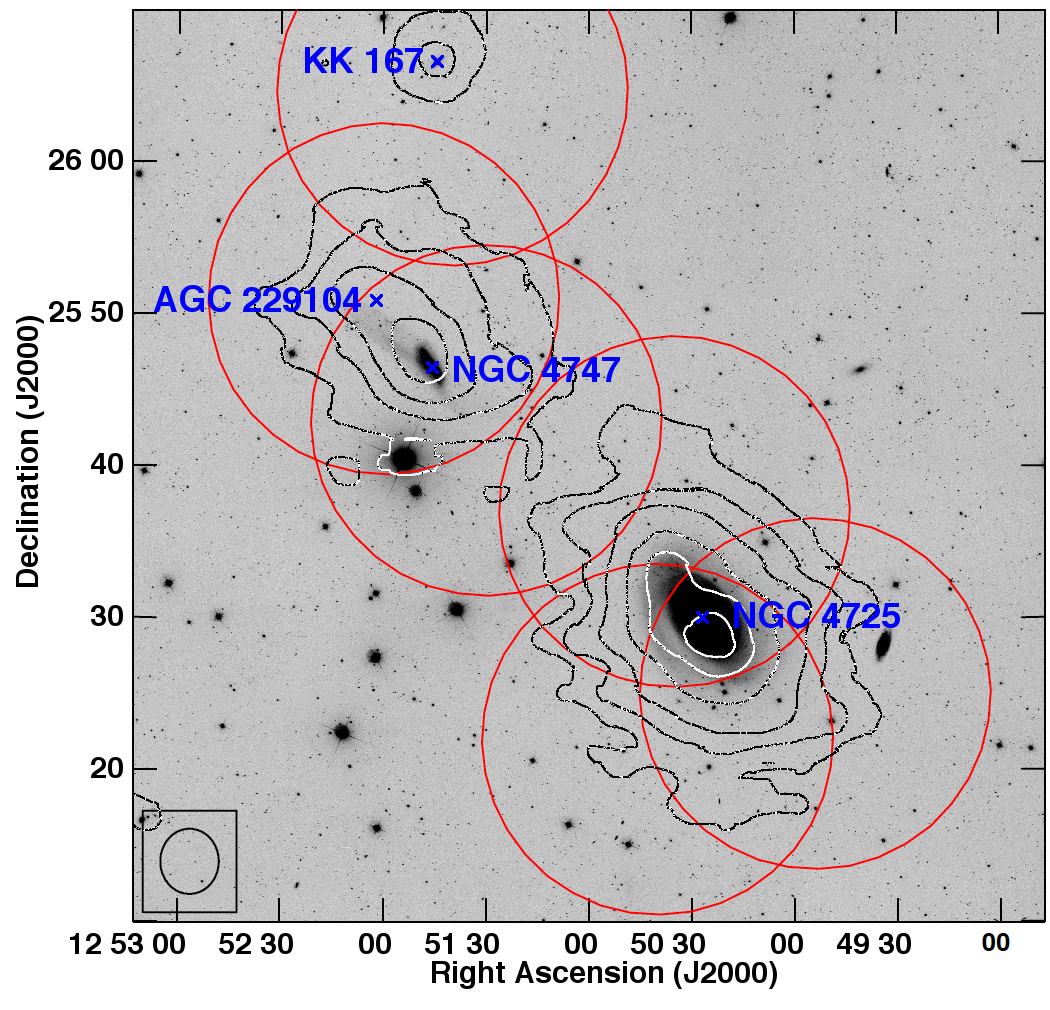}
  \caption[]{ALFALFA total {H\sc{i}} intensity contours at $N_{H_I} = (1.3, 4.0, 10.5, 25, 40, 50) \times 10^{19}$ atoms cm$^{-2}$ superimposed on a SDSS \textit{r}-band greyscale image of the NGC~4725/47 group.  The heliocentric velocity of the {H\sc{i}} contained within the lowest contour ranges from 990 to 1430 km s$^{-1}$.  The 4 arcmin (15 kpc) ALFALFA beam is in the bottom-left corner and the blue x's indicate the locations of the ALFALFA {H\sc{i}} detections.  The large red circles indicate the GMRT follow-up pointings. 
\label{fig:pointings}}
\end{center}
\end{figure}

NGC~4725/47 is a galaxy group that contains two spirals, NGC~4725 and NGC~4747 (previously observed in {H\sc{i}} by \citealt{hay1979} and \citealt{wev1984}), and at least one dIrr galaxy KK~167 \citep{kar1998}.  NGC~4725 has a SABab visual morphology; whereas, NGC~4747 is a nearly edge-on system with a SBcd morphology \citep{wev1984}.  The latter has a visibly perturbed disk, possibly three stellar tails --- two directed north-east (NE) and one south-west (SW; see Fig.~\ref{fig:pointings}) --- and an {H\sc{i}} tail extending away from NGC~4725, which strongly suggest that the two spirals are tidally interacting \citep{hay1979}.  ALFALFA detects an {H\sc{i}} peak, AGC~229104, that coincides with the NE stellar tails \citep{hay2011}.  The median distance to NGC~4725 --- derived from Cepheid variable stars --- is 13 Mpc \citep{gib1999}, which is the assumed distance for all group members throughout this paper.

This paper presents additional observational components of our multi-wavelength investigation of the gas-rich dwarf galaxy populations of three interacting groups.  In Section~\ref{gmrt}, we present high-resolution interferometric {H\sc{i}} observations from the GMRT and the gas properties of each detectable group member in NGC~4725/47.  Deep optical imaging from the CFHT MegaCam and estimates of the stellar properties for the low-mass objects found in NGC~3166/9 and NGC~4725/47 are detailed in Section~\ref{CFHT}.  Section~\ref{results} summarizes the results of the low-mass galaxy populations in the two groups.  In Section~\ref{discuss}, we present the culmination of our study of gas-rich dwarf galaxies and tidal features in NGC~3166/9, NGC~871/6/7 and NGC~4725/47, and compare our results to TDG simulations found in the literature.


\section{GMRT Observations of NGC~4725/47}
\label{gmrt}

Using the ALFALFA map as a pointing guide and following a similar observational set-up and reduction technique as described in Papers~1 and 2, the GMRT data consist of six pointings observed --- at a central frequency of 1414.8 MHz --- over seven nights in mid-December 2012, early February 2013 and mid-August 2013 (Fig.~\ref{fig:pointings}).  The $\sim$48 hours of usable telescope time, which included calibration observations on standard flux calibrators (3C147, 3C138 and 3C286) and nearby phase calibrators 1227+365 and 1330+251, were divided between the six pointings.  The observation set-up and map parameters are summarized in Table~\ref{param}.

\begin{table}
 \centering
 \begin{minipage}{80mm}
 \caption[]{GMRT observation set-up and map parameters}
 \label{param}
\begin{tabular}{ l c @{} c}  
\hline
Parameter 										& Value		& Units		\\ 
\hline 
Number of pointings								& 6			&			\\
Average time on source per pointing				& 330		& min		\\ 
Primary beam HPBW per pointing				& 19.7		& arcmin	\\
Mosaicked map size								& 80 		& arcmin	\\ 
Central observing frequency						& 1414.8 	& MHz		\\
Observing bandwidth							& 4.16		& MHz		\\ 
Observing spectral resolution						& 8.1		& kHz		\\ 
Final cube spectral resolution						& 24.4		& kHz		\\
Final cube spectral resolution						& 5.2		& km s$^{-1}$\\
Map spatial resolutions							& 30 and 45	& arcsec	\\ 
Peak map sensitivity:\\
\hspace{5mm}45 arcsec angular resolution			& 1.5		& mJy beam$^{-1}$\\  
\hspace{5mm}30 arcsec angular resolution			& 1.2		& mJy beam$^{-1}$\\  
Peak sensitivity around NGC~4747: \\ 
\hspace{5mm}45 arcsec angular resolution			& 1.8		& mJy beam$^{-1}$\\ 
\hspace{5mm}30 arcsec angular resolution			& 1.4		& mJy beam$^{-1}$\\ 
\hspace{5mm}15 arcsec angular resolution			& 1.2		& mJy beam$^{-1}$\\ 
\hline
\end{tabular} 
\end{minipage}
\end{table}

\begin{figure*}
\begin{center}
  \includegraphics[width=180mm]{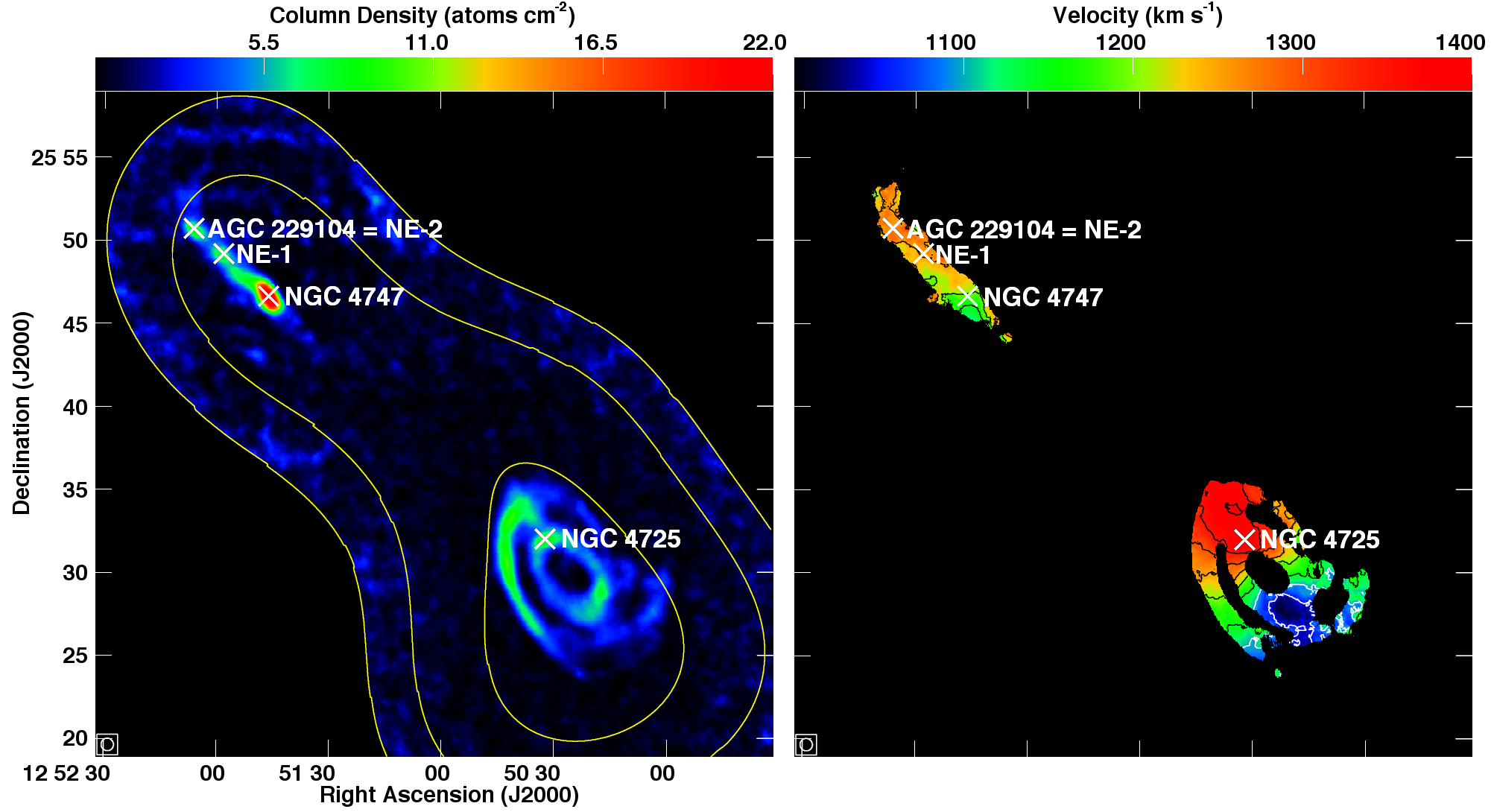}
  \caption[]{GMRT 45 arcsec angular resolution maps of the {H\sc{i}} in the central region of NGC~4725/47.  The x's indicate the central peak {H\sc{i}} flux density of the gas-rich detections (see Table~\ref{HImass}, column 2) and the synthesized beam is shown in the bottom-left corner of each panel.  Left panel: Total intensity (mom$_0$) map.  The yellow contours show the regions in the map with noise $\sigma = (1.6, 2, 2.5)$ mJy beam$^{-1}$ and the colour scale ranges linearly from 1 to $22 \times 10^{20}$ atoms cm$^{-2}$.  Right panel: Intensity-weighted velocity (mom$_1$) map.  Velocity contours are shown at 50 km s$^{-1}$ increments.  To remove spurious signals from both maps, regions in each channel where $\sigma > 2.5 $ mJy beam$^{-1}$ --- which includes the region around KK~167 --- were blanked before the moments were computed.  In addition, the velocity map was blanked at locations where the column density ($N_{H_I}$) $\leq 2\times10^{20}$ atoms cm$^{-2}$ and around the edges of the map.
\label{fig:GMRTgroup}}
\end{center}
\end{figure*}

\begin{figure*}
\begin{center}
  \includegraphics[width=150mm]{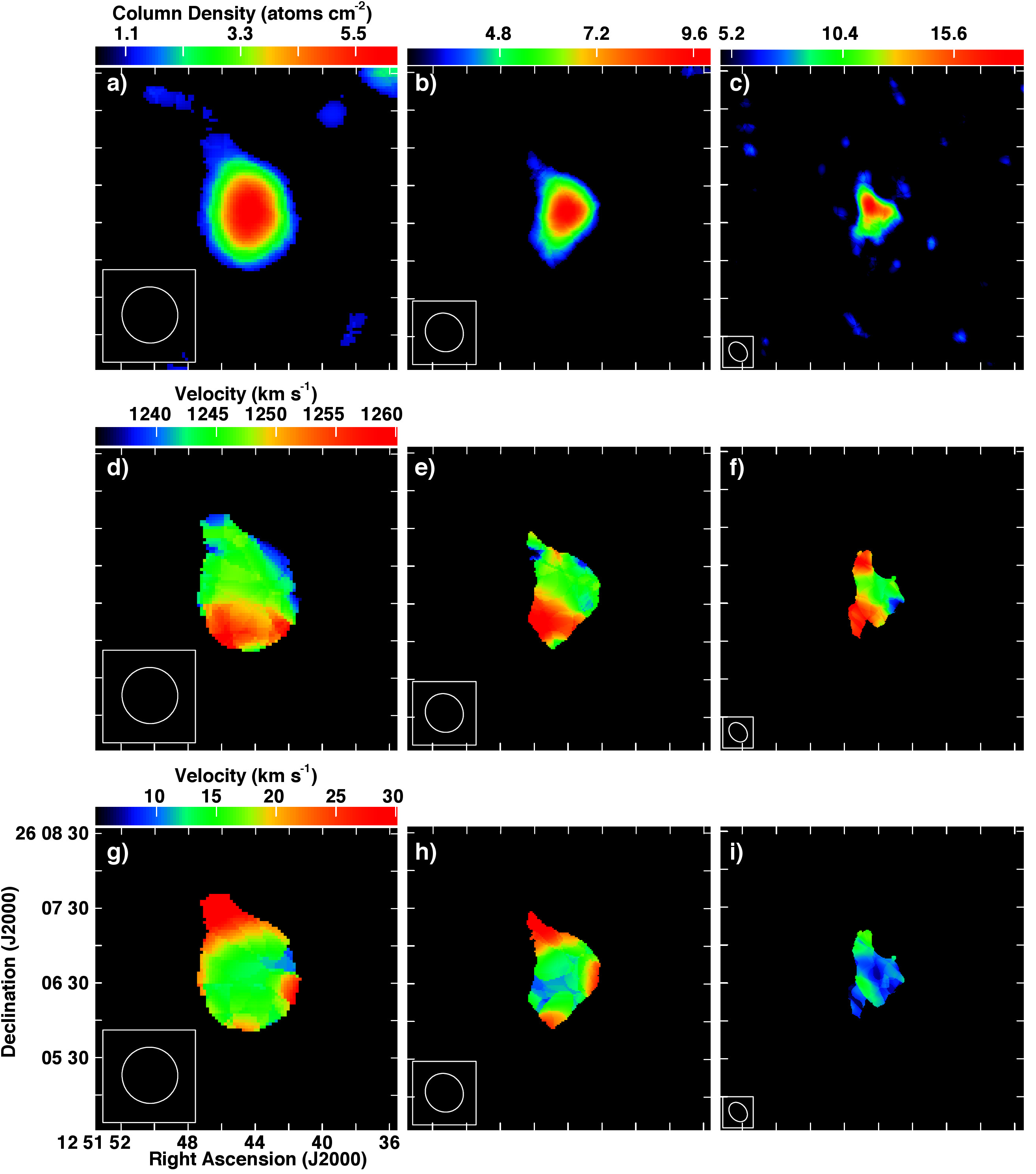}
  \caption[]{GMRT moment maps of KK~167.  Top row: mom$_0$ maps; middle row: mom$_1$ maps; bottom row: second velocity moment (mom$_2$) maps at various angular resolutions (first column: 45 arcsec; second column: 30 arcsec; third column: 15 arcsec).  The column density ($N_{H_I}$) ranges from (a) 0.55 to $6.3 \times 10^{20}$ atoms cm$^{-2}$ (b) 2.5 to $10 \times 10^{20}$ atoms cm$^{-2}$ (c) 4.7 to $19 \times 10^{20}$ atoms cm$^{-2}$.  The colour scales for the mom$_1$ and mom$_2$ maps range from 1235 to 1260 km s$^{-1}$ and 5 to 30 km s$^{-1}$, respectively, for all three angular resolutions.  To show the noise level for each resolution, no blanking was applied to the mom$_0$ maps; whereas, the mom$_1$ and mom$_2$ maps were blanked outside of the lowest contour level of the mom$_0$ maps.
  \label{fig:GMRTkk167}}
\end{center}
\end{figure*}

\begin{figure*}
\begin{center}
  \includegraphics[width=180mm]{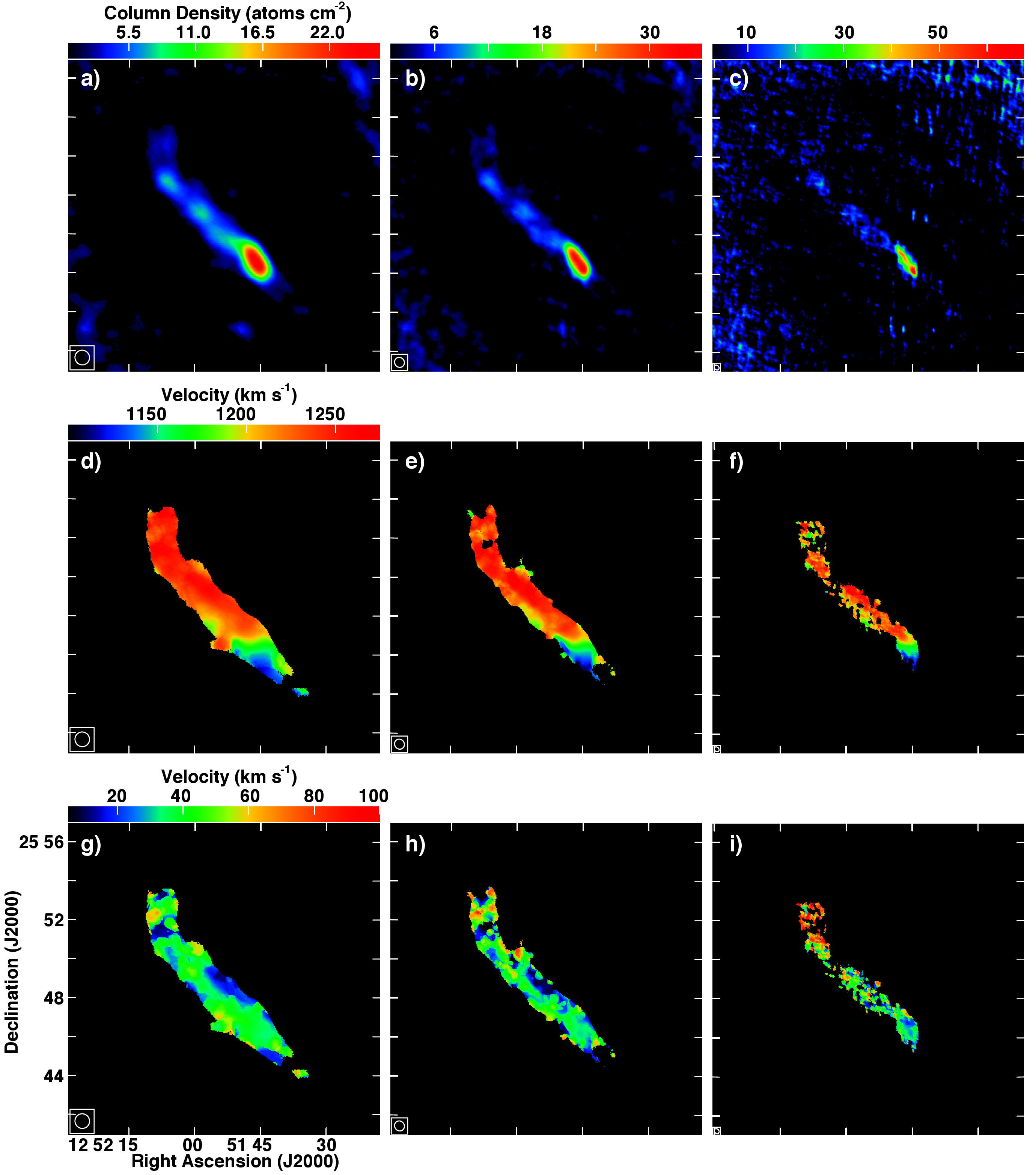}
  \caption[]{GMRT moment maps of NGC~4747 and its gas-rich tidal tail, which encompasses tidal knots NE-1 at 12 51 57.9, +25 49 14 and NE-2 (= AGC~229104) at 12 52 06.1, +25 50 42.  Top row: mom$_0$ maps; middle row: mom$_1$ maps; bottom row: mom$_2$ maps at various angular resolutions (first column: 45 arcsec; second column: 30 arcsec; third column: 15 arcsec).  $N_{H_I}$ ranges from (a) 0.55 to $26 \times 10^{20}$ atoms cm$^{-2}$ (b) 1.0 to $40 \times 10^{20}$ atoms cm$^{-2}$ (c) 2.5 to $68 \times 10^{20}$ atoms cm$^{-2}$.  The colour scales for the mom$_1$ and mom$_2$ maps range from 1100 to 1275 km s$^{-1}$ and 5 to 100 km s$^{-1}$, respectively, for all three angular resolutions.  Noise blanking for the mom$_1$ and mom$_2$ maps was applied in the same manner as described for Fig.~\ref{fig:GMRTkk167}.

\label{fig:GMRTn4747}}
\end{center}
\end{figure*}

Data editing and reduction were completed using the Astronomical Image Processing System (AIPS) version 31Dec14 \citep{gre2003} in the same manner as presented in Paper~1.  The calibrated data cubes were mosaicked and imaged with tapering to produce 45 arcsec (= 3 kpc at the distance of NGC~4725/47; see Fig.~\ref{fig:GMRTgroup}) and 30 arcsec (2 kpc) synthesized beam sizes for the group maps.  Higher resolution (and lower sensitivity) 15 arcsec (1 kpc) maps were also produced for the regions around KK~167 as well as NGC~4747 and its gas-rich tail (Figs \ref{fig:GMRTkk167} and \ref{fig:GMRTn4747}).  A three-channel average, resulting in a spectral resolution of 5.3 km s$^{-1}$ (complementary to ALFALFA's 5.2 km s$^{-1}$ cubes) was used to produce the final maps.  The pointing centred on KK~167 is significantly noisier ($\sigma = 4$ mJy beam$^{-1}$) than the others in the mosaic.  Accordingly, Fig.~\ref{fig:GMRTgroup} omits this pointing and shows only the central region of the group.  The multiple moment maps at various resolutions in Figs \ref{fig:GMRTkk167} and \ref{fig:GMRTn4747} provide a detailed view of the {H\sc{i}} in KK~167 and the gas-rich tidal tail of NGC~4747.

The GMRT observations are able to resolve the {H\sc{i}} in NGC~4725, KK~167, NGC~4747 and AGC~229104, which were previously detected by ALFALFA.  In addition, the interferometric maps distinguish an {H\sc{i}} knot, NE-1, between NGC~4747 and AGC~229104 (see Fig.~\ref{fig:GMRTn4747}).  Both NE-1 and AGC~229104 (hereafter NE-2) appear to be second-generation features formed from the gaseous material pulled out from NGC~4747 during an interaction event. 

\begin{figure}
\begin{center}
  \includegraphics[width=84mm]{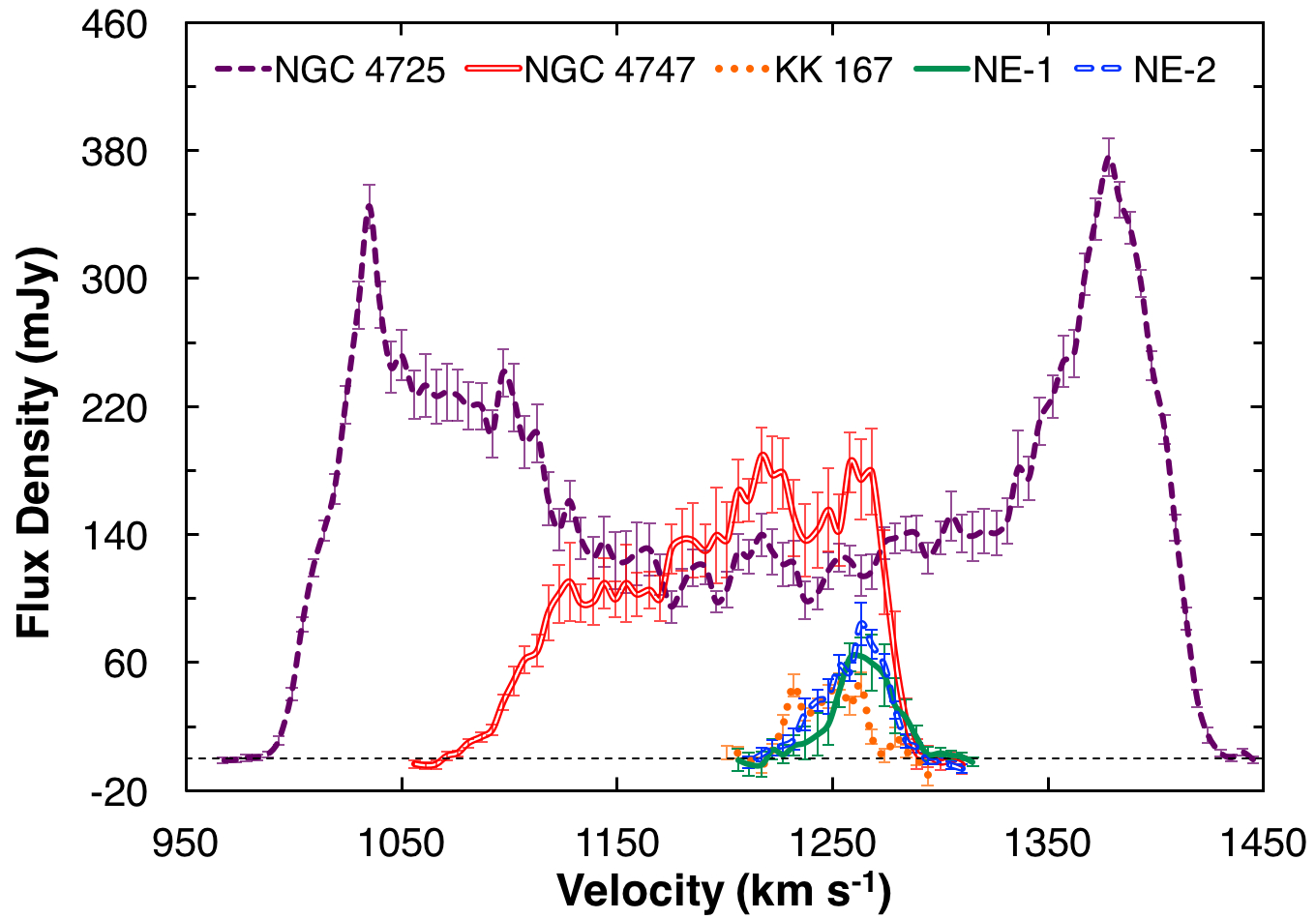}
  \caption[]{Global profiles for NGC~4725/47 group members derived from the GMRT 45 arcsec angular resolution data.
  \label{fig:globalprofile}}
\end{center}
\end{figure}

The flux density of each detection was measured from the higher sensitivity 45 arcsec angular resolution GMRT maps and used to produce the global profiles shown in Fig.~\ref{fig:globalprofile}.  The {H\sc{i}} mass can be calculated using:
\begin{equation}
M_{H_I} [M_{\odot}] = 2.356 \times 10^5 d^2 S_{H_I}
\end{equation}
where $d$ is the distance to the source in Mpc and $S_{H_I}$ is the flux density of the global profiles integrated over velocity in Jy km s$^{-1}$ \citep{gio1988}.  The {H\sc{i}} properties of each detection are presented in Table~\ref{HImass}.  The GMRT observations recover $\sim$80\% of the {H\sc{i}} mass measured by ALFALFA for the relatively isolated dIrr and two spiral galaxies; however, the GMRT recovers only $\sim$20\% of the ALFALFA-measured flux for NE-2; in fact, ALFALFA detects more than twice the {H\sc{i}} mass of the GMRT NE-1 and NE-2 detections combined.  It is apparent that much of the gas in the region surrounding the tidal knot is diffusely distributed on scales greater than $\sim$5 arcmin (20 kpc) and is resolved out by the GMRT.

\begin{table*}
\centering
 \begin{minipage}{180mm}
\caption[]{{H\sc{i}} detections in NGC~4725/47.  Column 1: detection name.  Column 2: right ascension and declination of central peak {H\sc{i}} flux density.  Column 3: centroid of the most probable optical counterpart.  Column 4: noise in region of detection in 45 arcsec GMRT maps.  Column 5: integrated {H\sc{i}} flux density computed from the global profiles in Fig.~\ref{fig:globalprofile}.  The uncertainty was determined by propagating the error from the global profile and adding a 20\% calibration error.  Column 6: {H\sc{i}} mass calculated from GMRT observations.  Column 7: {H\sc{i}} mass from ALFALFA $\alpha.40$ data release \citep{hay2011}.}
\label{HImass}
\begin{tabular}{l c c @{}c c c c} 
\hline
Name			&GMRT Coordinates		&Optical Coordinates	&$\sigma$			&$S_{H_I}$			&$M_{H_I}$(GMRT)		&$M_{H_I} (\alpha.40)$\\
				&(J2000) 				&(J2000)				&(mJy beam$^{-1}$)	&(Jy km s$^{-1}$)	&($10^8 M_{\odot}$)		&($10^8 M_{\odot}$)\\
(1)				&(2) 					& (3) 					& (4)				&(5)				&(6)					&(7)\\	
\hline 
NGC~4725		&12 50 32.2, +25 32 00	&12 50 26.6, +25 30 03	& 1.6				&70 $\pm$ 20		& 30 $\pm$ 6 			&46.20 $\pm$ 0.08\\
KK~167			&12 51 44.2, +26 06 36	&12 51 44.3, +26 06 38	& 4.1				&2 $\pm$ 1	 		& 0.7 $\pm$ 0.4 			&0.96 $\pm$ 0.02\\
NGC~4747 		&12 51 46.2, +25 46 38	&12 51 45.7, +25 46 28	& 1.8				&24 $\pm$ 5 		& 9 $\pm$ 2 			&11.37 $\pm$ 0.04\\
NE-1			&12 51 57.9, +25 49 14 	&--						& 1.9				&2.2 $\pm$ 0.7 		& 0.9 $\pm$ 0.3 			&--\\
NE-2 (= AGC~229104)	&12 52 06.1, +25 50 42	&--				 & 2.0				&2.7 $\pm$ 0.7 		& 1.1 $\pm$ 0.3 			&4.94 $\pm$ 0.02\\
\hline 
\end{tabular}
\end{minipage}
\end{table*}

Assuming that each gas-rich group member is self-gravitating and in dynamical equilibrium, the total dynamical mass of each object can be computed using:
\begin{equation}
M_{dyn} [M_{\odot}] = 3.39 \times 10^4 a_{H_I} d \left( \frac{W_{20}}{2\sin i} \right)^2
\end{equation}
where $a_{H_I}$ is the {H\sc{i}} major axis diameter of the object in arcmin, $d$ is the distance to the source in Mpc, $W_{20}$ is the velocity width --- at 20\% of the peak flux density --- of each object in km s$^{-1}$ and $i$ is the reported inclination.  
$a_{H_I}$ is measured from the $N_{H_I} = 1 \times 10^{20}$ atom cm$^{-2}$ contour of the 30 arcsec angular resolution GMRT maps (the lowest statistically significant contour for all measured sources) and corrected for beam smearing effects.

The dynamical information for each detection is in Table~\ref{dyn}.  The uncertainty in $M_{dyn}$ is computed as the average half-difference between the masses computed using the more conservative $W_{20}$ value and other velocity width estimates (i.e.~using the conventional $W_{50}$ --- velocity width at 50\% of the peak flux density --- or the velocity difference across $a_{H_I}$ of each source).  Note that NE-1 and NE-2 do not have velocity gradients that are consistent with rotation (see Fig.~\ref{fig:GMRTn4747}).  It is possible that either of these objects might be face-on {H\sc{i}} disks, where an inclination correction would have significant affect on $M_{dyn}$, but it is more likely that they are dominated by non-circular motions (i.e. pressure supported).  The computed values of $M_{dyn}$ for these features adopt $\sin i = 1$ and have meaning only if they are in dynamical equilibrium.

\begin{table}
\centering
 \begin{minipage}{84mm}
\caption[]{Dynamical information for NGC~4725/47 group members.  Column 1: detection name.  Column 2: global profile width measured at 20\% of the peak flux density.  Column 3: inclination from $^{a}$HyperLeda database \citep{pat2003} and $^{b}$Karachentsev Catalogue \citep{kar1998}, where applicable.  Column 4: beam-corrected major axis diameter of the object measured from the 30 arcsec resolution maps.  Column 5: heliocentric velocity of the profile midpoint at 20\% of the peak flux density.  Column 6: total dynamical mass estimated using Equation 5.2, where $\sin i$ = 1 for objects with no specified inclinations (see text for further details).}
\label{dyn}
\begin{tabular}{l c @{}c @{} c c c} 
\hline
Source 			& $W_{20}$ 	&$i$	 		&$a_{H_I}$			&$cz_{\odot}$	&$M_{dyn}$ \\
				& (km s$^{-1}$)	&(deg)			&(arcmin)			&(km s$^{-1}$)	&($10^8 M_{\odot}$) \\
(1)				& (2) 			& (3)			&(4)				&(5)			&(6) \\
\hline
NGC~4725 		& 412 $\pm$ 5 	& 45$^{a}$		& 13.4 $\pm$ 0.5	& 1209 $\pm$ 3	& 5100 $\pm$ 500\\
KK 167			& 46 $\pm$ 5 	& 45$^{b}$		& 1.6 $\pm$ 0.4		& 1246 $\pm$ 3	& 7 $\pm$ 1\\
NGC~4747		& 184  $\pm$ 5 	& 64$^{a}$  		& 3 $\pm$ 1			& 1190 $\pm$ 3	& 150 $\pm$ 20\\
NE-1			& 49 $\pm$ 5 	&--				& 1.6 $\pm$ 0.4		& 1265 $\pm$ 3	& 4 $\pm$ 1\\
NE-2			& 48 $\pm$ 5 	&--				& 1.9 $\pm$ 0.6		& 1258 $\pm$ 3	& 5 $\pm$ 1\\
\hline
\end{tabular}
\end{minipage} 
\end{table}


\section{CFHT Observations}
\label{CFHT}

The CFHT MegaCam, with a 1~deg$^2$ field of view, was used to obtained deep $g'r'i'$-band imaging of the groups in this study.  Similar to the set-up described in Paper~2, a seven-point large dithering pattern (LDP-CCD-7), with exposures sequenced within a 1 hour time-frame to ensure minimal sky variations, was used for each filter.  There was sufficient background in each image to allow for sky modelling and subtraction using the Elixir/Elixir-LSB processing pipeline (\citealt{mag2004}, \citealt{cui2011}), negating the requirement for off-target frames. The photometric images were stacked using Elixir to characterize and subtract the background and then processed through Elixir-LSB to remove the scattered light components (see \citealt{detal2011}, \citealt{fer2012}).  The final image pixels were binned $3\times3$ ($0.56 \times 0.56$ arcsec$^{2}$) to boost the signal-to-noise of the optical features.  The zeropoint for each band is ZP = 29.61 mag and physical structures observed in the images can be detected down to $\sim$1 ADU above the sky background, which is equivalent to a detection limit of 28.4 mag arcsec$^{-2}$ across the three bands.

Observations in $g'$-band for NGC~4725/47 (centred on NGC~4747, see below) and in all three bands for NGC~3166/9 were completed in semester 2013A.  The remaining $r'$- and $i'$-band imaging of NGC~4725/47 were completed in semester 2014A.  Since NGC~4725/47 spatially spans $\sim$1 deg (230 kpc) two sets of LDP-CCD-7 exposures --- one centred on NGC~4725 and the other on NGC~4747 --- were conducted in each of the two observing bands.  The images were processed and stacked for each field separately and then stitched together for each band, which introduced faint artifacts in the final images.  Since the stitched image does not significantly improve the sensitivity in the regions where the images overlap, the individual images centred on NGC~4747 were used for the measurements in this paper.  The set-up parameters and resulting image properties for the utilized fields are summarized in Table~\ref{cfht_n4725}. 

\begin{table}
 \centering
 \begin{minipage}{84mm}
 \caption[]{CFHT observation set-up and image properties}
 \label{cfht_n4725}
\begin{tabular}{ l c c c}  
\hline
Parameter 										& $g'$-band		& $r'$-band		&$i'$-band	\\ 
\hline 
\textbf{NGC~4725/47:}							&				&				&\\
Exposure time (sec)								& 345  			& 345			& 270		\\ 
Number of exposures							& 7				& 7				& 7		\\
Mean image quality (arcsec)						& 1.16			& 1.16			& 1.19		\\
Sky background, $3\times3$ bin (ADU)			& 387.2			& 768.6			& 1187.2	\\
Sky brightness (mag arcsec$^{-2}$)				& 22.0			& 21.8			& 20.7		\\
Detection limit (mag arcsec$^{-2}$)				& 28.4			& 28.4			& 28.4		\\ 
\hline
\textbf{NGC~3166/9:}							&				&				&\\
Exposure time (sec)								& 345  			& 345			& 270		\\ 
Number of exposures							& 7				& 7				& 7		\\
Mean image quality (arcsec)						& 0.88			& 0.74			& 0.97		\\
Sky background, $3\times3$ bin (ADU)			&463.9			& 778.6			&1284.0		\\
Sky brightness (mag arcsec$^{-2}$)				& 21.8			& 21.8			& 20.6		\\
Detection limit (mag arcsec$^{-2}$)				& 28.4			& 28.4			& 28.4		\\ 
\hline
\end{tabular} 
\end{minipage}
\end{table}

\begin{figure*}
\begin{center}
    \includegraphics[width=180mm]{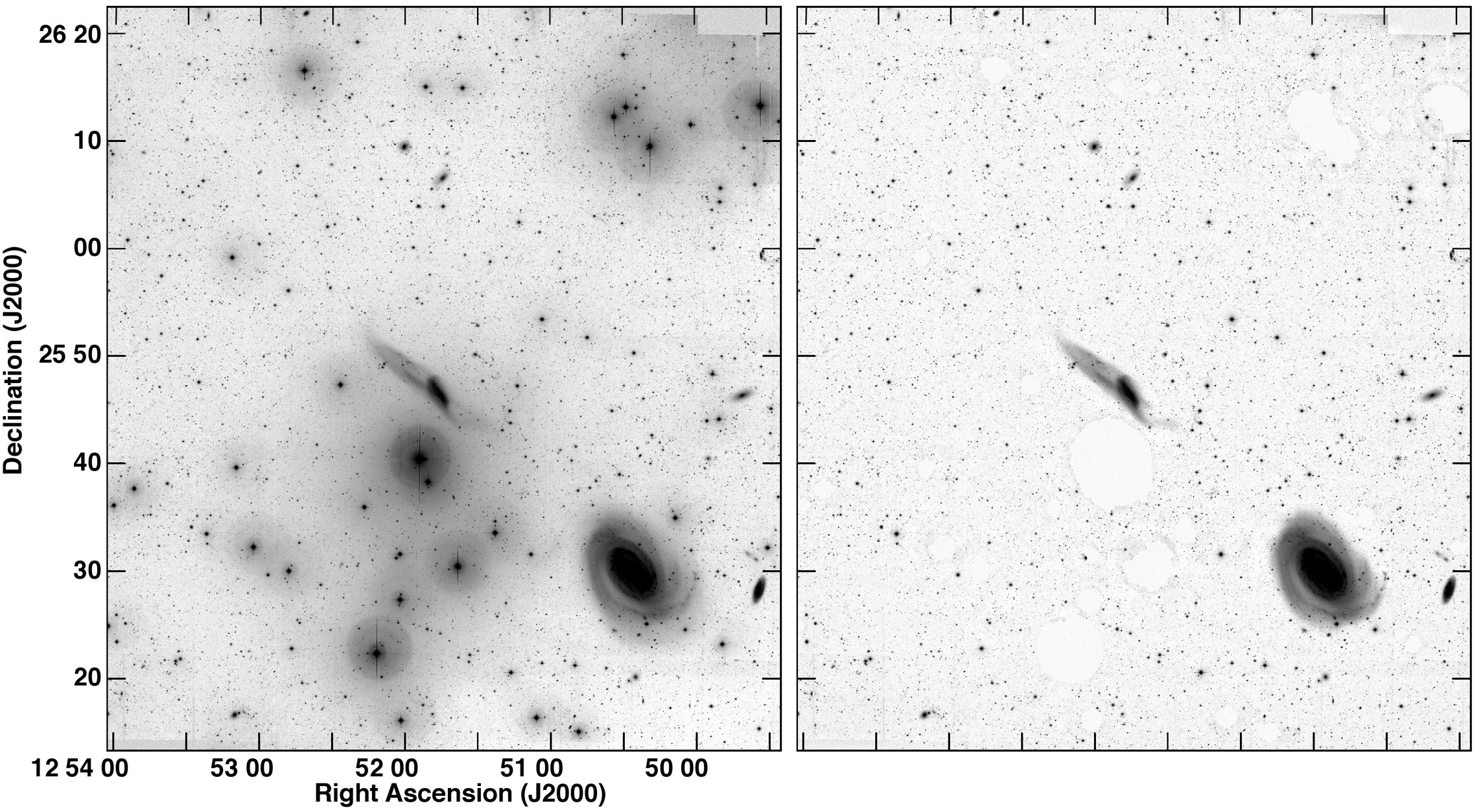}
  \caption[]{CFHT MegaCam $g'$-band image of the NGC~4725/47 field.  Left panel: image produced by the Elixir/Elixir-LSB processing pipeline.  Right panel: the same field with bright foreground stars, their reflection halos and other artifacts subtracted.
  \label{fig:subtract}}
\end{center}
\end{figure*}

\begin{figure*}
\begin{center}
  \includegraphics[trim=0 -20mm 0 0, width=140mm]{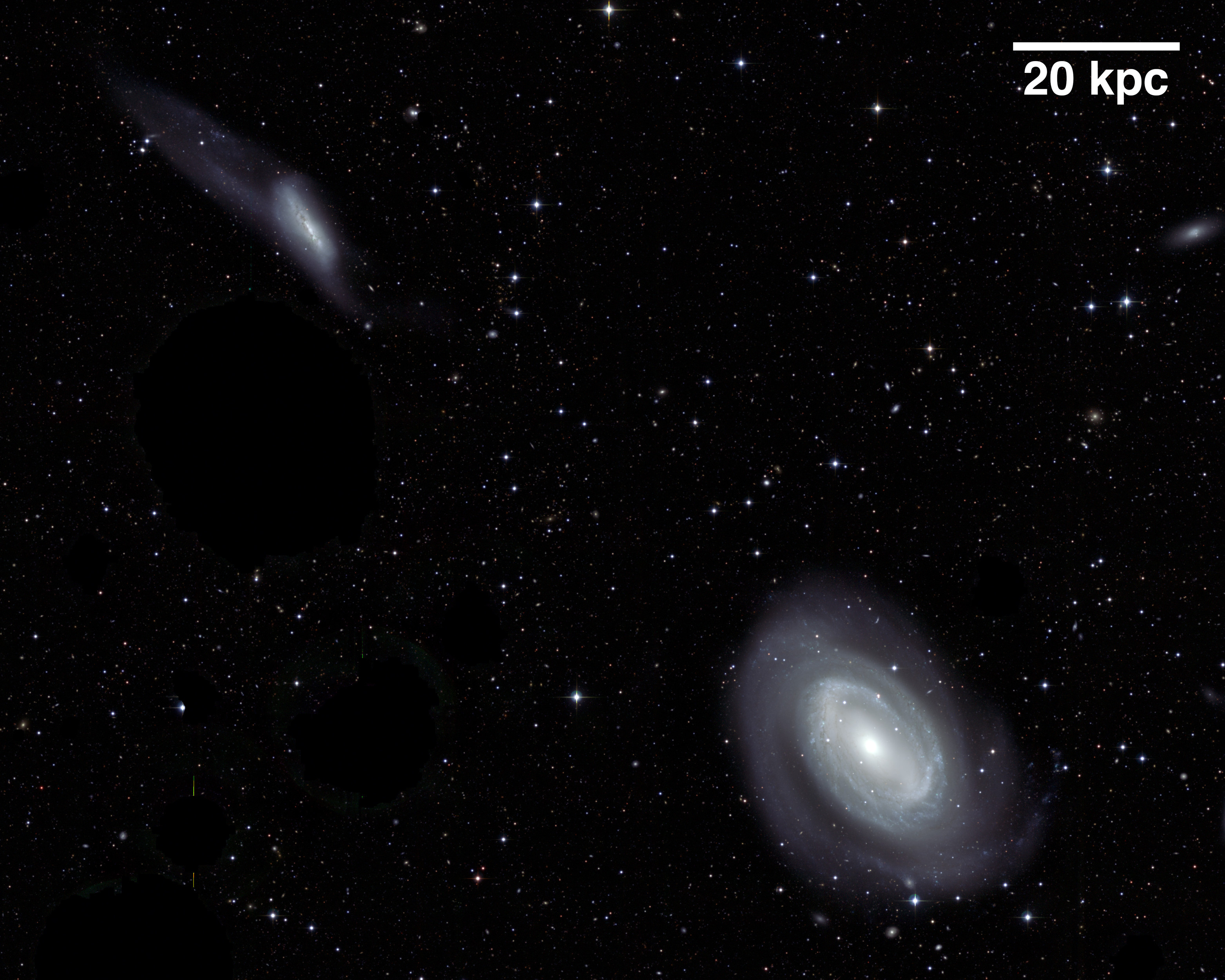}
    \includegraphics[width=140mm]{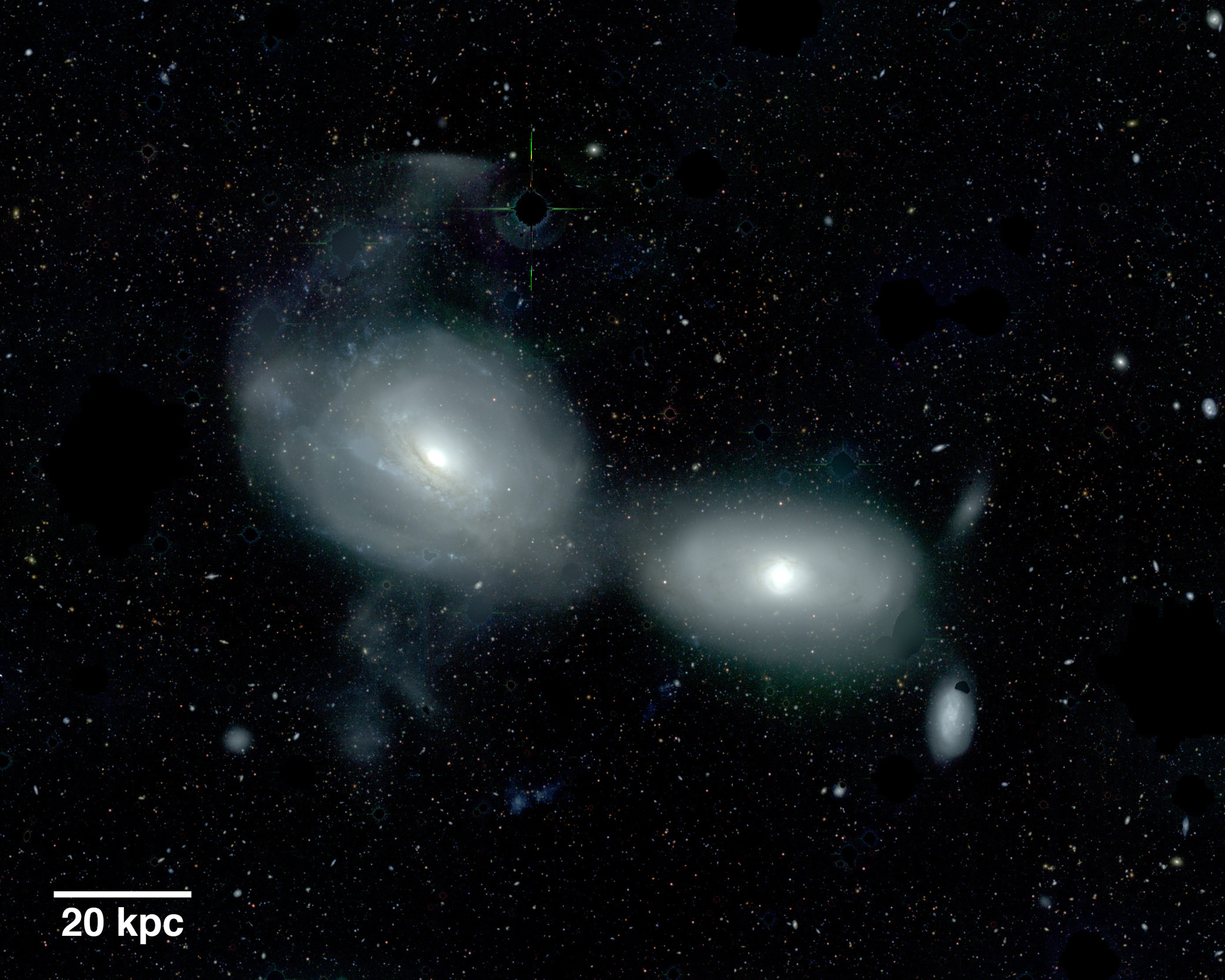}
  \caption[]{CFHT MegaCam $g'r'i'$-band composite images.  The horizontal white bar is 20 kpc across at the distance of each group.  Top: NGC~4725/47.  Bottom: NGC~3166/9.  Note: interpolations of surrounding areas have been used to fill-in any blank regions --- resulting from the subtraction of foreground stars --- near the key features in the group.
  \label{fig:CFHT}}
\end{center}
\end{figure*}

For both sets of photometric maps, bright foreground stars and their resultant reflection halos could not be excised by Elixir/Elixir-LSB and left artifacts in the optical images (see the left panel of Fig.~\ref{fig:subtract}).  After additional background subtraction (using grid sampling measurements and interpolations to compute the background behind galaxies and key features) in each band, these bright stars and associated halos were modelled using a variable centre ring technique (see \citealt{wan2010}) and then subtracted from the images.  This grid sampling method produced a lattice of 7.5 $\times$ 7.5 arcmin regions that was utilized for local background estimates.  After visual inspection, stellar spikes, saturated parts of stars and other noticeable artifacts were manually removed (see the right panel of Fig.~\ref{fig:subtract}).  Any subtracted regions spatially coinciding with key features were filled-in using interpolations of the surrounding areas.  Fig.~\ref{fig:CFHT} shows the fully processed $g'r'i'$-band composite images for the central regions of NGC~4725/47 and NGC~3166/9.  

\subsection{Optical and Ancillary Data Analysis: NGC~4725/47}
\label{4725}

\begin{figure*}
\begin{center}
  \includegraphics[width=180mm]{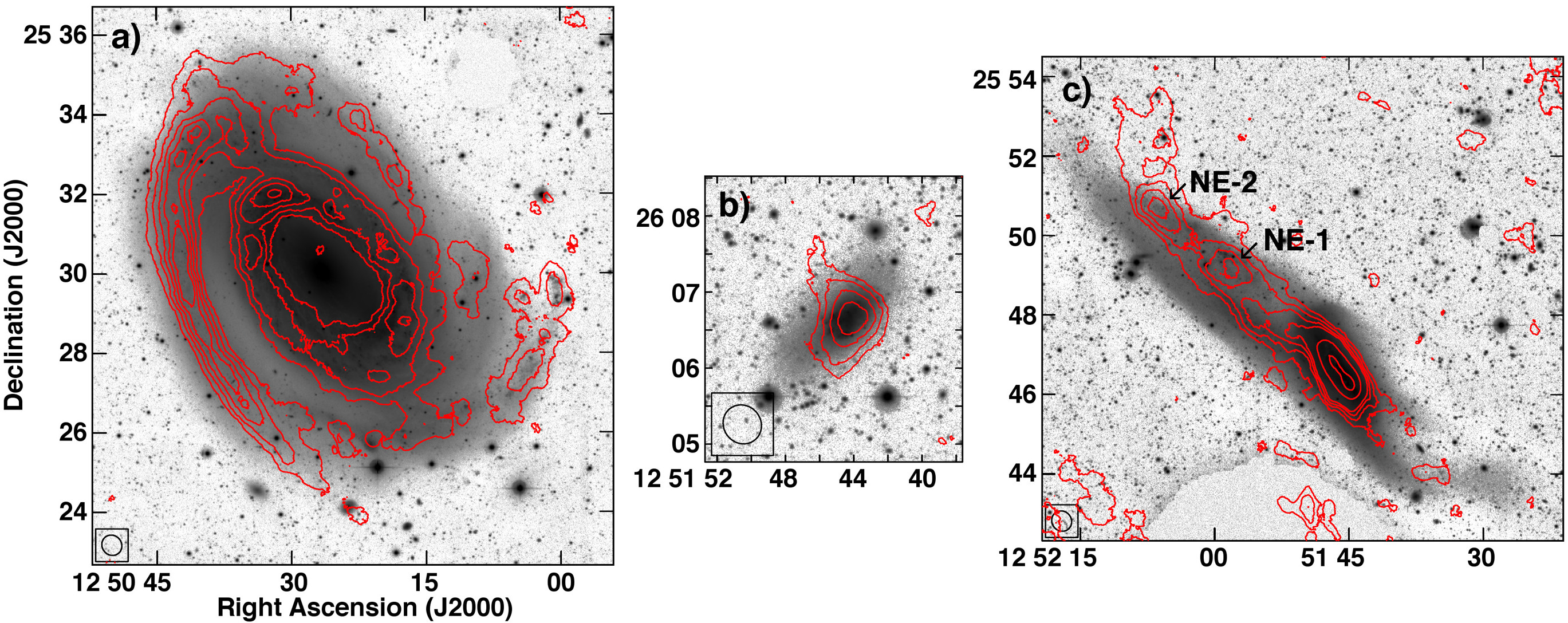}
  \caption[]{GMRT 30 arcsec angular resolution mom$_0$ contours superimposed on CFHT $g'$-band images.  a) NGC 4725.  The GMRT mom$_0$ contours are at $N_{H_I}$ = (2, 4, 6, 8, 10) $\times 10^{20}$ atoms cm$^{-2}$.  b) KK~167.  The contours are at $N_{H_I}$ = (2.5, 4, 6, 8) $\times 10^{20}$ atoms cm$^{-2}$.  c) NGC 4747 and its tidal tails.  The contours are at $N_{H_I}$ = (2, 4, 6, 8, 19, 32) $\times 10^{20}$ atoms cm$^{-2}$.  
  \label{fig:NGC4725_cfht}}
\end{center}
\end{figure*}

GMRT mom$_0$ contours superimposed on CFHT $g'$-band images, to show the spatial coincidence of gaseous and stellar components of the gas-rich group members, are presented in Fig.~\ref{fig:NGC4725_cfht}.  There are two stellar knots --- Stellar knot - 1 is located at 12 51 58.8, +25 49 33 (originally identified by \citealt{wev1984}) and the other fainter feature, Stellar knot - 2 is at 12 52 02.5, +25 50 13 --- which appear to be within the NE stellar tail; however, these knots are offset from the {H\sc{i}} peaks, which could indicate a different evolutionary history for the stellar and gaseous tidal tails.  It is also possible that the stellar knots are not physically associated with the tail (cf.~\citealt{mir1991}).

\begin{figure*}
\begin{center}
  \includegraphics[width=140mm]{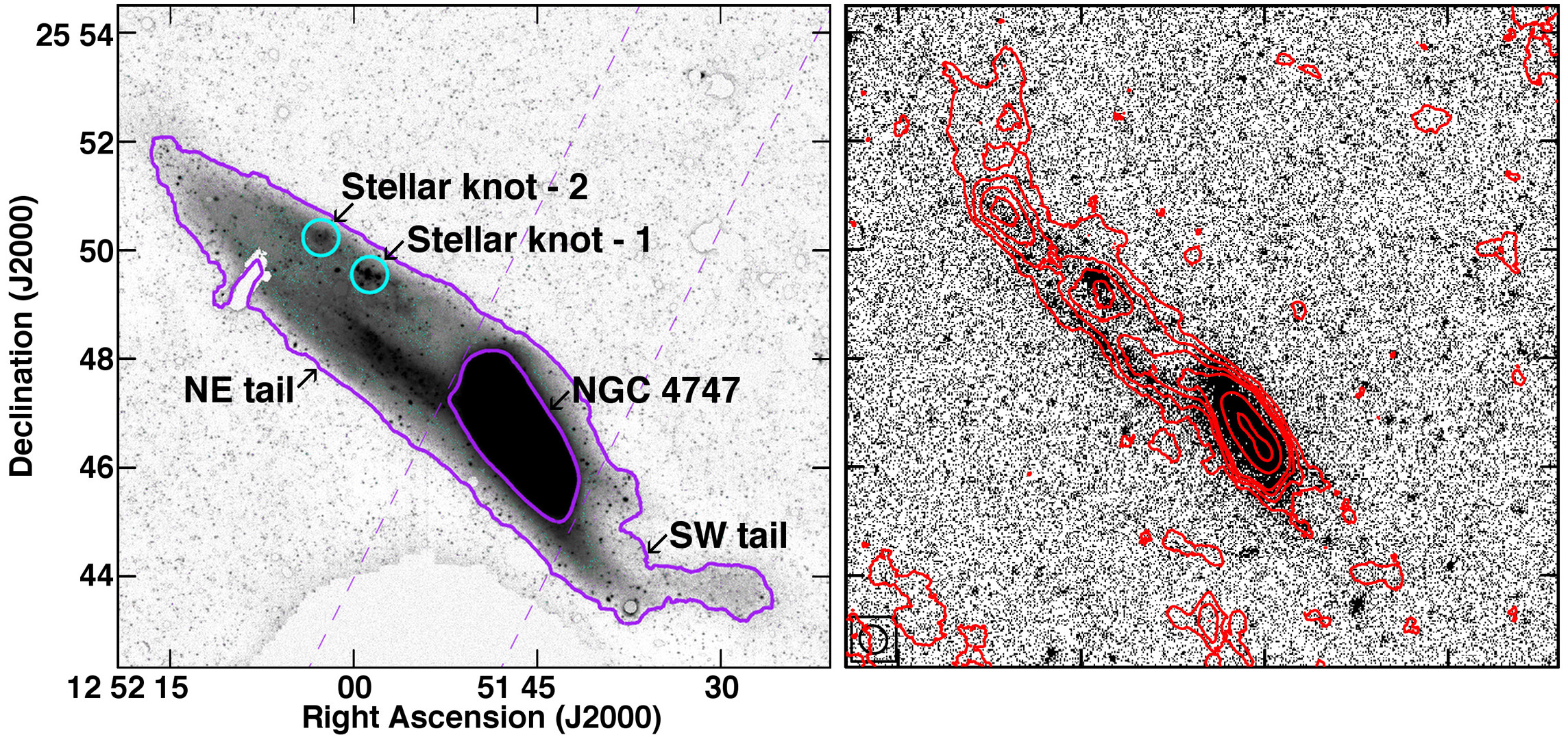}
  \caption[]{Left panel: CFHT $g'$-band image of NGC~4747 and its surrounding stellar tails.  The solid outlines enclose the regions that were used to measure the stellar properties of each object listed in Table~\ref{stellar}.  The dashed lines shows selected cutoff for each tail.  Right panel: \textit{GALEX} FUV image of the same region.  The GMRT mom$_0$ contours are at the same column densities as Fig.~\ref{fig:NGC4725_cfht}c.
  \label{fig:N4747extract}}
\end{center}
\end{figure*}

The optical magnitudes of NGC~4725 and KK~167 were measured using straight-forward aperture photometry.  Detailed deblending was required to separate NGC 4747 from its stellar tails.  The left panel of Fig.~\ref{fig:N4747extract} shows the boundary of the individually measured features.  For NE-1 and NE-2, their values were measured as the optical regions within the $5\times10^{20}$ atoms cm$^{-2}$ contour from the GMRT 30 arcsec angular resolution mom$_0$ maps.  The CFHT MegaCam magnitudes were then converted to more conventional SDSS $gri$ AB magnitudes using standard equations provided by the CFHT.  

The right panel of Fig.~\ref{fig:N4747extract} shows the GMRT contours superimposed on a far-ultraviolet (FUV) image from the \textit{Galaxy Evolution Explorer} (\textit{GALEX}), obtained from the GR6/GR7 data release.  It appears that the FUV tail aligns with the outer edge of the NE {H\sc{i}} tail, which is evidence for recent star formation activity (see Section~\ref{age_metal}).  Stellar knot - 1 is quite bright in the FUV and its location (near an {H\sc{i}} peak and situated along the FUV tail) suggests that it is probably a star-forming region within the tail, which warrants future follow-up optical spectroscopy to confirm its distance.  The optical and FUV images also indicate that what was once believed to be two stellar tails extending NE are likely a single bent tail (see \citealt{bri2004}).  This projection effect could imply an overestimate in the {H\sc{i}} masses presented in Table~\ref{HImass} for NE-1 and NE-2, which will be further discussed in Section~\ref{results:NGC4725/47}. 

In the same manner as presented in Paper~2, stellar masses, $M_{stellar}$, were computed using:
\begin{equation}
log~M_{stellar}/[M_{\odot}] = 1.15 + 0.7(g\mbox{-}i) - 0.4 \mbox{M}_i
\end{equation}
where M$_i$ is the absolute magnitude of each object in the $i$-band \citep{tay2011}.   This relation has been calibrated using an extensive census of galaxies with $M_{stellar} \geq 10^{7.5}$ $M_{\odot}$.  The optical properties of the group members and features in NGC~4725/47 are listed in Table~\ref{stellar}.  Since NE-1 and NE-2 do not appear to have bound stellar counterparts, their masses are estimated as upper limits.

\begin{table*}
 \centering
  \begin{minipage}{180mm}
\caption[]{Optical properties of detections in NGC~4725/47.  Column 1: detection name.  Column 2: radius defined by the 25 mag arcsec$^{2}$ isophote.  The irregular shape of the tidal tails as well as their proximity to NGC~4747 precludes accurate measurements of the radii of these objects.  Column 3: radius containing 50\% of the light.  Column 4: radius containing 90\% of the light.  Column 5: apparent magnitude in $g$-band.  Regions defining the stellar tails and knot are shown in the left panel of Fig.~\ref{fig:N4747extract}.  Column 6: $g$-$r$ colour.  Column 7: $g$-$i$ colour.  Column 8: absolute magnitude in $i$-band.  Column 9: stellar mass computed using Equation 5.3.}
\label{stellar}
\begin{tabular}{ l c c c c c c c c} 
\hline
Source 			&$r_{25}$	& r$_{50}$	&r$_{90}$	&$g$					&$g$-$r$					&$g$-$i$					&M$_{i}$			&$M_{stellar}$\\
				& (arcsec)	& (arcsec)	&(arcsec)	&(mag)					&(mag)					&(mag)					&(mag)				&($\times 10^8 M_{\odot}$)\\
(1)				& (2) 		& (3)		&(4)		&(5)					&(6)					&(7)					&(8)				&(9)\\
\hline
NGC~4725		& 376.4		&201.5		&473.0		&9.6809 $\pm$ 0.0001	&0.697 $\pm$ 0.009		&1.001 $\pm$ 0.009		&-21.9 $\pm$ 0.8	&$400\pm20$ \\ 
KK~167			& 31.1		&33.1		&108.1		&15.543 $\pm$ 0.003	&0.388 $\pm$ 0.006		&0.546 $\pm$ 0.006		&-15.6 $\pm$ 0.6	& $0.58\pm0.02$\\
NGC~4747		&--			&78.0		&145.8		&12.6389 $\pm$ 0.0004	&0.519 $\pm$ 0.008		&0.769 $\pm$ 0.005		&-18.7 $\pm$ 0.7	& $14.7\pm0.6$ \\
NE-1			&--			&--			&--			&16.223 $\pm$ 0.003	&0.35 $\pm$ 0.03		&0.49 $\pm$ 0.03		&-14.8 $\pm$ 0.6 	& $\leq0.3$ \\
NE-2			&--			&--			&--			&17.155 $\pm$ 0.005	&0.35 $\pm$ 0.01		&0.50 $\pm$ 0.01		&-13.9 $\pm$ 0.5 	& $\leq0.2$ \\
SW tail 			&--			&--			&--			&15.986 $\pm$ 0.003	&0.53 $\pm$ 0.03		&0.84 $\pm$ 0.03		&-15.4 $\pm$ 0.6 	& $0.81\pm0.04$ \\
NE tail			&--			&--			&--			&13.994 $\pm$ 0.001	&0.443 $\pm$ 0.004		&0.634 $\pm$ 0.004		&-17.2 $\pm$ 0.7	& $3.0\pm0.1$ \\
Stellar knot - 1	&--			&--			&--			&17.232 $\pm$ 0.004	&0.305 $\pm$ 0.005		&0.423 $\pm$ 0.007		&-13.8 $\pm$ 0.5	& $0.089\pm0.004$ \\
Stellar knot - 2	&--			&--			&--			&17.752 $\pm$ 0.006	&0.308 $\pm$ 0.008		&0.46 $\pm$ 0.01		&-13.3 $\pm$ 0.5	& $0.061\pm0.003$ \\
\hline
\end{tabular}
\end{minipage}
\end{table*}


\subsection{Optical Analysis: NGC~3166/9}
\label{3166}

\begin{figure}
\begin{center}
  \includegraphics[width=84mm]{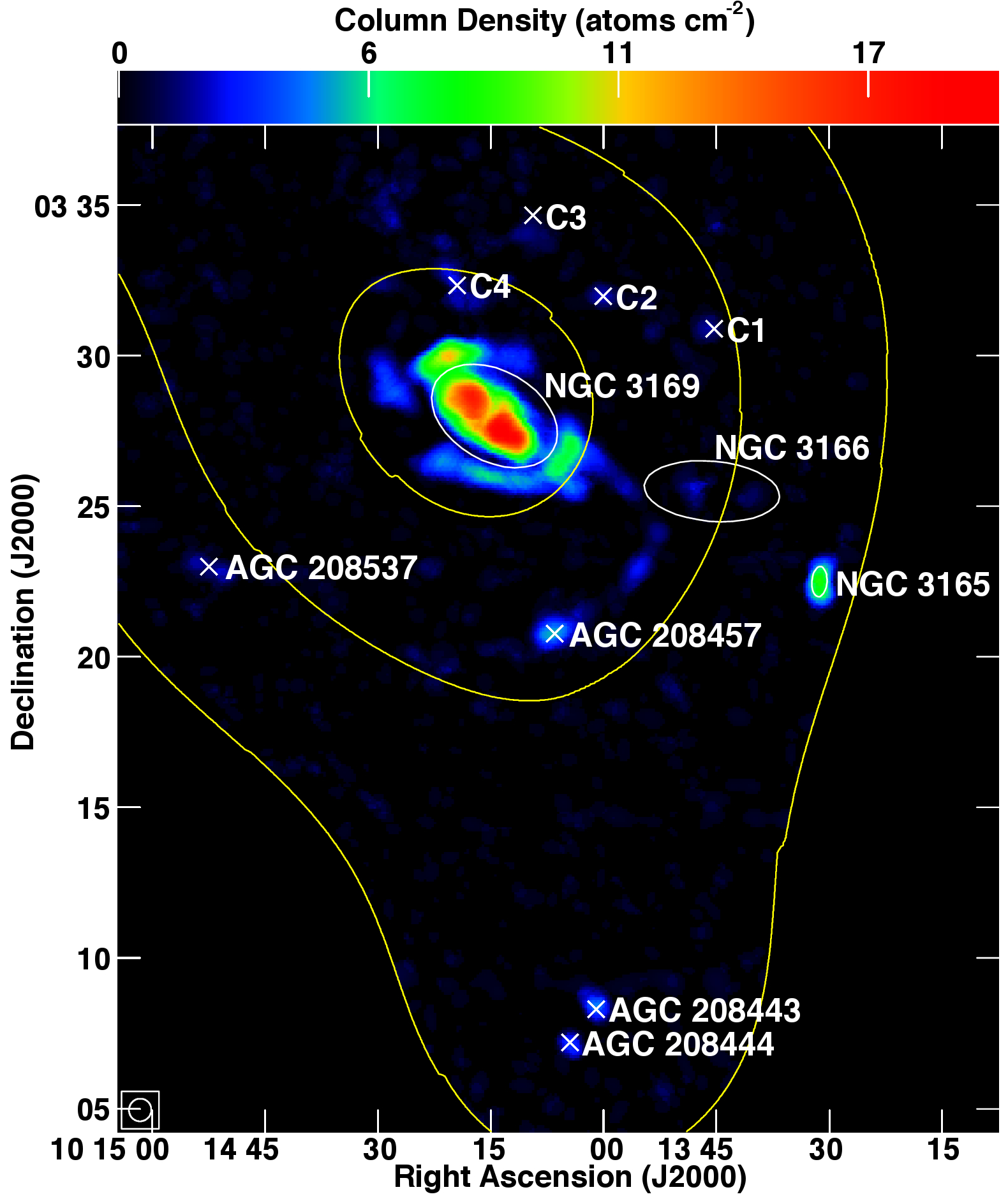}
  \caption{GMRT 45 arcsec angular resolution mom$_0$ map of the {H\sc{i}} in NGC 3166/9 (reproduced from Paper 1).  The ellipses represent the optical counterparts of the NGC galaxies and the x's indicate the locations of the low-mass objects.  The yellow contours show the regions in the map with map noise $\sigma = (1.7, 2, 2.8)$ mJy beam$^{-1}$.  \label{fig:n3166_group}}
\end{center}
\end{figure}

As originally discussed in Paper~1, eight low-mass {H\sc{i}} objects were detected in the GMRT observations of NGC~3166/9.  An {H\sc{i}} map of this group is shown in Fig.~\ref{fig:n3166_group} and preliminary properties of the group members are presented in Table~\ref{n3166_lowmass}.  Based on our measurements in that paper, AGC~208537, AGC~208443 and AGC~208444 are likely classical dIrrs; the latter two galaxies appear to be interacting with each other while falling in towards the central region of the group.  AGC~208457 is a putative TDG that has the hallmarks of a self-gravitating, dark matter poor, second-generation galaxy.  The initial analysis of the properties of  {H\sc{i}} knots C1 to C4 could not conclusively distinguish these object as either dIrrs or tidal knots, although their masses and locations imply the latter interpretation (see Paper~1 for further details).  

\begin{figure}
\begin{center}
  \includegraphics[width=84mm]{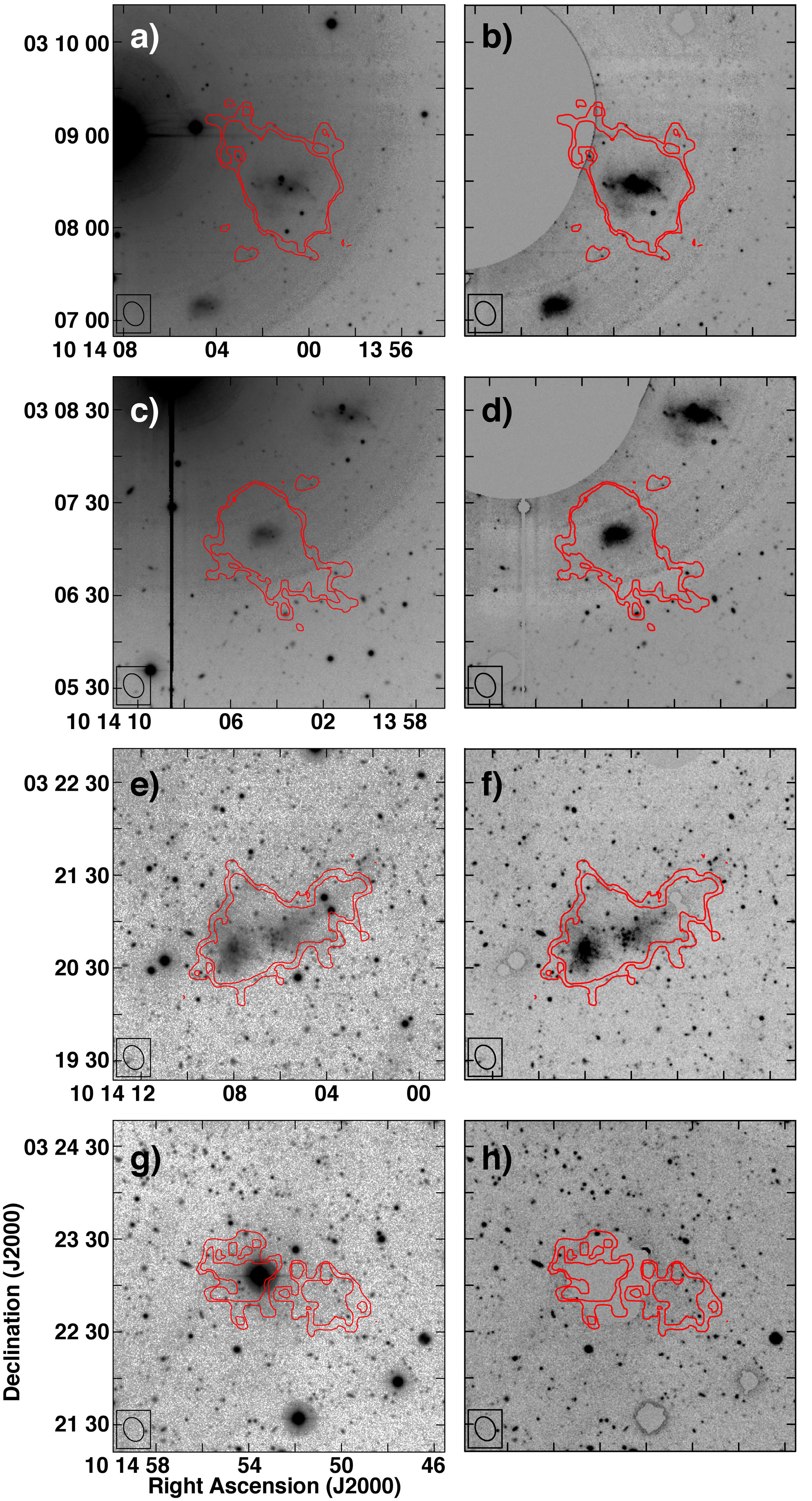}
  \caption[]{GMRT 15 arcsec angular resolution mom$_0$ contours at $N_{H_I}$ = (2.5, 3.75) $\times 10^{19}$ atoms cm$^{-2}$ of the dIrrs and TDG candidate in NGC~3166/9 superimposed on CFHT $g'$-band images, which have been scaled to highlight possible stellar counterparts of each group member.  Left panels: images produced by the Elixir/Elixir-LSB pipeline.  Right panels: same regions with bright foreground stars and other artifacts subtracted.  a,b) AGC~208443.  In order to preserve the optical counterparts of AGC~208443 and AGC~208444, the reflection halo of a nearby star, HD 88725, could not be fully subtracted.  c,d) AGC~208444.  e,f) AGC~208457 (TDG candidate).  g,h) AGC~208537.  
  \label{fig:lowmassNGC3166_1}}
\end{center}
\end{figure}

\begin{figure}
\begin{center}
  \includegraphics[width=84mm]{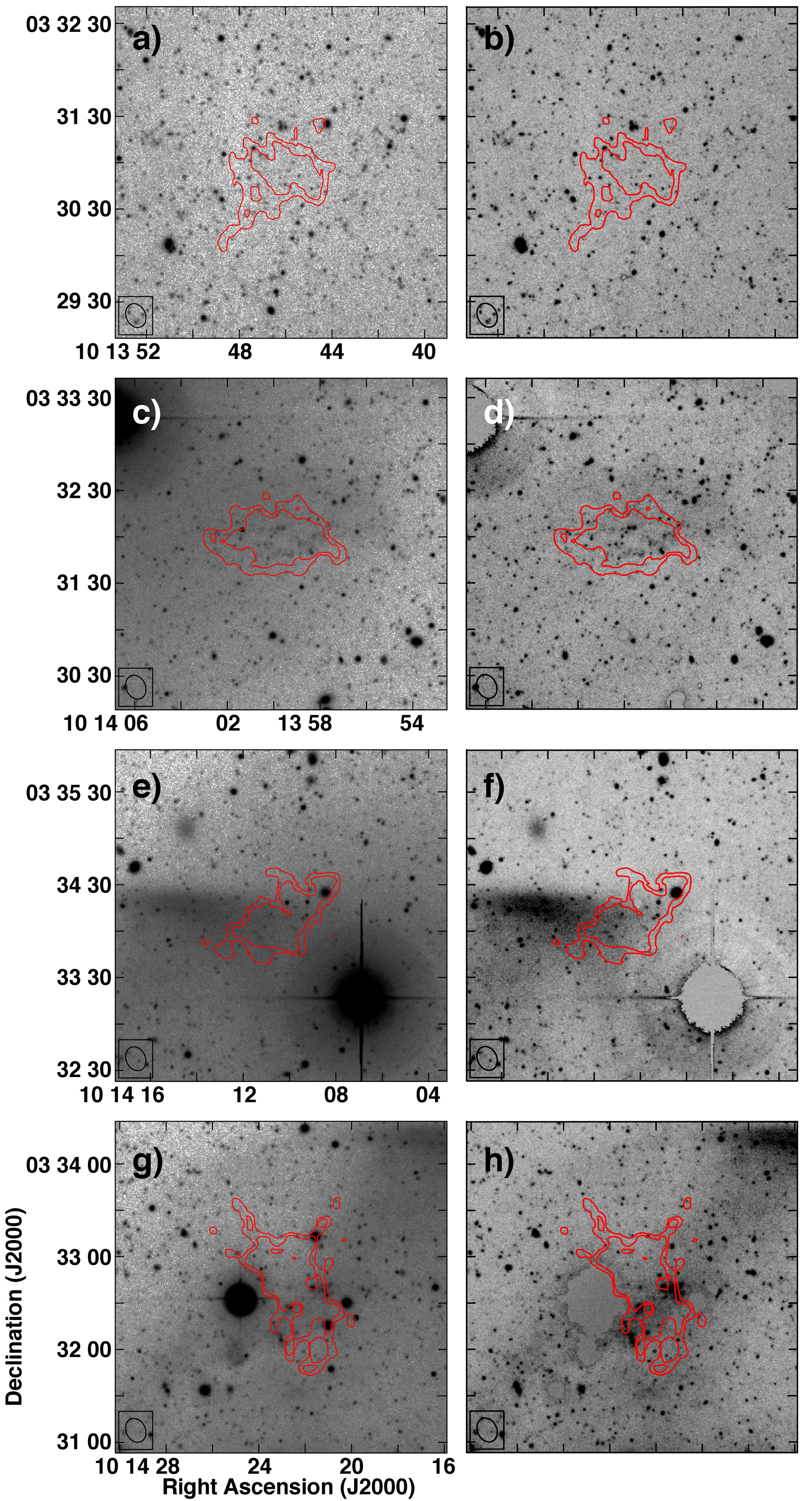}
  \caption[]{GMRT 15 arcsec angular resolution mom$_0$ of the low-mass {H\sc{i}} knots in NGC~3166/9 superimposed on CFHT $g'$-band images.  The contours are at the same level as in Fig.~\ref{fig:lowmassNGC3166_1}.  Left panels: Elixir+Elixir-LSB images.  Right panels: same regions with bright foreground stars and other artifacts subtracted.  a,b) C1.  c,d) C2.  e,f) C3.  g,h) C4.
    \label{fig:lowmassNGC3166_2}}
\end{center}
\end{figure}

GMRT mom$_0$ contours superimposed on CFHT $g'$-band images, which are scaled to highlight any optical counterparts of the low-mass {H\sc{i}} detections, are shown in Fig.~\ref{fig:lowmassNGC3166_1} (dIrrs and TDG candidate) and Fig.~\ref{fig:lowmassNGC3166_2} ({H\sc{i}} knots C1 to C4).  The reflection halo from a nearby bright star, HD 88725 (at 10 14 08, +03 09 04), could not be fully subtracted while preserving the optical counterparts of AGC~208443 and AGC~208444 and has left artifacts in the images (see Fig.~\ref{fig:lowmassNGC3166_1}a-b).  Aperture photometry was utilized to measure the group members with spatially coincident stellar features.  In order to minimize the flux contribution of the residual reflection halo around AGC~208443 and AGC~208444, apertures were manually fit for these two objects.

\begin{table*}
\centering
 \begin{minipage}{115mm}
\caption[]{Gas-rich group members in NGC~3166/9.  Column 1: detection name.  Column 2: right ascension and declination of peak {H\sc{i}} flux density.  Column 3: heliocentric velocity.  Column 4: {H\sc{i}} mass (from Paper~1 GMRT observations).  Column 5: morphological classification notes from Paper~1.}
\label{n3166_lowmass}
\begin{tabular}{l c c c c} 
\hline
Source 				&GMRT Coordinates		&$cz_{\odot}$		&$M_{H_I}$				&Classification\\
					&(J2000) 				&(km s$^{-1}$)		&($10^8 M_{\odot}$)		&Notes\\
(1)					&(2)					&(3) 				&(4)					&(5)\\	
\hline
NGC~3165 			&10 13 31.0, +03 22 30	& 1324$\pm$ 3 		&1.8 $\pm$ 0.3 			& spiral: SAdm\\
NGC~3166			&10 13 41.4, +03 24 44	& 1326$\pm$ 4		&--						& spiral: SAB0\\
C1 					&10 13 45.3, +03 30 53	& 1302 $\pm$ 3		& 0.31 $\pm$ 0.08		& {H\sc{i}} knot\\ 
C2 					&10 14 00.0, +03 31 58	& 1307 $\pm$ 3		& 0.5 $\pm$ 0.1			& {H\sc{i}} knot\\ 
AGC~208443 		&10 14 01.0, +03 08 18 	& 1482 $\pm$ 3		& 1.0 $\pm$ 0.2 			& in-falling dIrr\\ 
AGC~208444 		&10 14 04.5, +03 07 13 	& 1488 $\pm$ 3		& 0.7 $\pm$ 0.1			& in-falling dIrr\\ 
AGC~208457 		&10 14 06.5, +03 20 47	& 1343 $\pm$ 3		& 2.3 $\pm$ 0.3 			& TDG candidate\\ 
C3 					&10 14 10.0, +03 34 10	& 1110 $\pm$ 3		& 0.5 $\pm$ 0.1 			& {H\sc{i}} knot\\ 
NGC~3169 			&10 14 15.0, +03 28 00	& 1248$\pm$ 3 		& 42 $\pm$ 5 			& spiral: SAa\\
C4 					&10 14 20.4, +03 32 40	& 1028 $\pm$ 3		& 0.35 $\pm$ 0.08 		& {H\sc{i}} knot\\ 
AGC~208537 		&10 14 52.5, +03 23 00 	& 1182 $\pm$ 3		& 0.6 $\pm$ 0.1 			& dIrr\\ 
\hline 
\end{tabular}
\end{minipage} 
\end{table*}

Even in the deep CFHT optical images, the low-mass {H\sc{i}} knots in NGC~3166/9 --- i.e.~C1 to C4 --- do not have obvious bound optical counterparts (Fig.~\ref{fig:lowmassNGC3166_2}).  There does seem to be an extremely faint stellar over-density that spatially coincides with C2 (see Fig.~\ref{fig:lowmassNGC3166_2}b) and C3 and C4 appear to lie on this stellar tail around NGC~3169, which likely has tidal origins; however, the magnitudes for the {H\sc{i}} knots were measured as the optical region within the $1\times 10^{20}$ atoms cm$^{-2}$ contour of the GMRT 15 arcsec mom$_0$ maps.  The stellar masses for these knots are given as upper limits in Table~\ref{n3166_stellar}, which presents the optical properties (with CFHT MegaCam magnitudes converted into SDSS $gri$ AB magnitudes) of the group members.  Note: AGC~208537 has been omitted from Table~\ref{n3166_stellar} as its {H\sc{i}} spatially coincides with a subtracted star, SDSS J101453.53+032305.9, which prohibits accurate measurements of an optical component (see Fig.~\ref{fig:lowmassNGC3166_1}h).

\begin{table*}
 \centering
 \begin{minipage}{175mm}
\caption[]{Optical properties of gas-rich group members in NGC~3166/9.  Column labels are the same as those in Table~\ref{stellar}.  AGC~208537 has been omitted from this table as its {H\sc{i}} spatially coincides with a subtracted star.  *The magnitude and colour measurements for AGC~208443 and AGC~208444 are likely affected by the residual halo of a foreground star (see Section~\ref{3166}).  These values have been included for completeness.}
\label{n3166_stellar}
\begin{tabular}{ l c c c c c c c c} 
\hline
Source 			&$r_{25}$	& r$_{50}$	&r$_{90}$	&$g$					&$g$-$r$					&$g$-$i$					&M$_{i}$			&$M_{stellar}$\\
				& (arcsec)	& (arcsec)	&(arcsec)	&(mag)					&(mag)					&(mag)					&(mag)				&($\times 10^8 M_{\odot}$)\\
(1)				& (2) 		& (3)		&(4)		&(5)					&(6)					&(7)					&(8)				&(9)\\
\hline
NGC~3165		&42.7		&16.6		&34.1		&14.2514 $\pm$ 0.0009	&0.472 $\pm$ 0.003		&0.705 $\pm$ 0.008		&-18.2 $\pm$ 0.2	&8.6 $\pm$ 0.1	\\ 
NGC~3166		&162.7		&26.4		&107.2		&10.8653 $\pm$ 0.0002	&0.744 $\pm$ 0.009		&1.143 $\pm$ 0.009		&-22.1 $\pm$ 0.3	&590 $\pm$ 10\\ 
C1 				&--			&--			&--			&20.1 $\pm$ 0.3			&1.4 $\pm$ 0.5			&1.3 $\pm$ 0.5			&-13.0 $\pm$ 0.3	&$\leq 0.2 $\\ 
C2 				&--			&--			&--			&18.08 $\pm$ 0.06		&0.4 $\pm$ 0.2			&0.3 $\pm$ 0.1			&-14.0 $\pm$ 0.2	&$\leq 0.09 $\\ 
AGC~208443* 	&16.4		&11.2		&22.8		&16.69 $\pm$ 0.03		&-0.22 $\pm$ 0.03		&0.10 $\pm$ 0.03		&-15.2 $\pm$ 0.2	&0.20 $\pm$ 0.06\\ 
AGC~208444*	&11.4		&8.9		&17.0		&17.57 $\pm$ 0.02		&0.47 $\pm$ 0.03		&0.39 $\pm$ 0.04		&-14.6 $\pm$ 0.2	&0.18 $\pm$ 0.02\\ 
AGC~208457	&--			&23.9		&47.8		&17.95 $\pm$ 0.04		&0.28 $\pm$ 0.06		&0.49 $\pm$ 0.05		&-14.3 $\pm$ 0.2	&0.16 $\pm$ 0.02\\ 
C3 				&--			&--			&--			&17.9 $\pm$ 0.1			&1.4 $\pm$ 0.2			&1.4 $\pm$ 0.2			&-15.2 $\pm$ 0.2	&$\leq 1.7 $\\ 
NGC~3169		&208.1		&50.7		&149.0		&10.7038 $\pm$ 0.0002	&0.758 $\pm$ 0.002		&1.126 $\pm$ 0.002		&-22.2 $\pm$ 0.3	&650 $\pm$ 10\\  
C4 				&--			&--			&--			&17.292 $\pm$ 0.007	&0.613 $\pm$ 0.009		&0.93 $\pm$ 0.02		&-15.4 $\pm$ 0.2	&$\leq 0.9 $\\ 
\hline
\end{tabular}
\end{minipage}
\end{table*}


\subsection{Stellar Population Analysis}
\label{age_metal}

Following a similar method to that utilized in Paper~2 to estimate the age of the optical counterpart to each group member, $g$-$r$ and $g$-$i$ colours were compared to the \citet{bc2003} stellar population models. These models  assume two separate initial mass functions (IMFs; i.e.~\citealt{cha2003} and \citealt{sal1955}) and use various spectral libraries covering a range of ages ($10^5$ to $10^{10}$ yr) and metallicities (0.0001 $< Z <$ 0.05).  Ideally, the pair of colours for each detection would correspond to a single population of stars; however, without detailed metallicity information to constrain the stellar models, significant degeneracies exist.  The range of ages reported for each feature encompasses all the models that are consistent with the $g$-$r$ and $g$-$i$ colours in the metallicity range given above.

Table~\ref{final} summarizes the dynamical and stellar properties of the group members and features in NGC~4725/47 and NGC~3166/9.  We note the large range of ages possible for NGC~4747 given its colours.  This range reflects the strong dependence of age on metallicity for this object.  For example, if this galaxy has a metallicity of $Z$ = 0.0001, its age would be 11.8 - 12.0 Gyr for a Salpeter IMF; however, if its metallicity is closer to solar ($Z_{\odot}$ = 0.02), then its age would be between $1.1 - 1.4$ Gyr.  The constraints on the other group members are somewhat better, allowing us to gain some insight by comparing their relative ages (see Section~\ref{results}).

Several low-mass object in NGC~3166/9 did not fit the stellar models due to contamination issues from foreground stars (i.e.~ACG~208443, AGC~208444 and AGC~208537) or because we are working with upper limits from undetected optical counterparts (i.e. C1 and C3).  The total gas mass, $M_{gas}$, for each detection was computed by multiplying $M_{H_I}$ by a factor of 1.33 to account for helium and other elements.  This value was added to $M_{stellar}$ to find the total baryonic mass, $M_{baryon}$, for each group member.  

We note that, although $M_{gas}$ for classical dIrrs is generally dominated by atomic gas, TDGs have been shown to contain an average molecular gas mass of $M_{mol} \sim 0.2 M_{H_I}$ \citep{bra2001}.  More evolved TDGs appear to have higher $M_{mol}/M_{H_I}$ ratios than those embedded within tidal tails \citep{bra2001}.  Since the {H\sc{i}} knots in NGC 3166/9 and NGC 4725/47 appear to be associated with tidal tails, the molecular gas in these objects is likely low, $M_{mol} \lesssim 0.1 M_{H_I}$, which is comparable to the uncertainties for $M_{gas}$.  We therefore do not include a contribution from molecular gas in the $M_{gas}$ estimates.

All dynamical masses have been computed under the assumption that each detection is self-gravitating in dynamical equilibrium, which is probably inaccurate for some of the objects as indicated by the values in brackets (see Section~\ref{results} for further details).  For the gas-rich objects with no clearly visible bound optical component, upper limits were estimated by measuring the background surface brightness level and/or diffuse stellar region within the $10^{20}$ atoms cm$^{-2}$ {H\sc{i}} contours. 

Similar to Papers 1 and 2, the luminosity --- $L_{FUV}$ in erg s$^{-1}$ Hz$^{-1}$ --- for each group member was measured from archival \textit{GALEX} FUV data to estimate SFRs using:
\begin{equation}
\mbox{SFR$_{FUV}$[$M_{\odot}$ yr$^{-1}$]} = 1.27 \times 10^{-28} L_{FUV}
\end{equation}
(\citealt{ken1998}, \citealt{hun2010}).  No additional corrections have been made to account for dust.  Gas-rich objects with no discernible FUV counterpart have upper limit SFR estimates based on the background noise.

\begin{table*}
 \centering
 \begin{minipage}{180mm}
\caption[]{Properties of the detected group members and features in NGC~4725/47 and NGC~3166/9.  Column 1: detection name.  Column 2: gas mass = 1.33$M_{H_I}$.  Column 3: total baryonic mass = $M_{gas} + M_{stellar}$.  AGC~208537 spatially coincides with a subtracted star, which prevents the measurement of a stellar mass and other optical properties.  Column 4: dynamical to gas mass ratio.  Note: NE-1, NE-2 and C1 to C4 are unlikely to be self-gravitating and therefore the computed values of $M_{dyn}$ may have little meaning.  These values have been include for completeness, but are indicated by the square brackets.  Column 5: dynamical to baryonic mass ratio.  Column 6: dynamical mass to $g$-band light ratio.  Column 7: apparent age of the stellar component (where applicable), assuming a Chabrier IMF.  The stellar ages have been determined within the systematic uncertainties of the \citet{bc2003} models.  Column 8: apparent age of the stellar component, assuming a Salpeter IMF.  Column 9: star formation rate for each object estimated using archival $GALEX$ FUV images.}
\label{final}
\begin{tabular}{l c c c c c c c c} 
\hline
Source 			& $M_{gas}$ 		&$M_{baryon}$		&$M_{dyn}/M_{gas}$ 	&$M_{dyn}/M_{baryon}$ 	&$M_{dyn}/L_g$		&Age - IMF$_C$	&Age - IMF$_S$		&SFR$_{FUV}$ \\
		& ($\times 10^8 M_{\odot}$)	&($\times 10^8 M_{\odot}$)	&				&				&($M_{\odot}/L_{\odot}$)		&($\times10^9$ yr)	&($\times10^9$ yr)	&($M_{\odot}$ yr$^{-1}$)\\
(1)				& (2) 				& (3)				&(4)					&(5)					&(6)				&(7)				&(8)				&(9)\\
\hline
\multicolumn{8}{l}{\textbf{NGC~4725/47:}}\\
NGC~4725		& 39 $\pm$ 8		& 440 $\pm$ 20		& 130 $\pm$ 30 			& 11 $\pm$ 1			& 26 $\pm$ 3		& 13.3 - 13.8		& 9.5 - 9.8			& 0.58 $\pm$ 0.02\\ 
KK~167			& 0.9 $\pm$ 0.5		& 1.5 $\pm$ 0.6		& 8 $\pm$ 4				& 5 $\pm$ 2				& 8 $\pm$ 1 		& 0.9 - 3.5			& 0.64 - 1.4 			& 0.010 $\pm$ 0.001\\ 
NGC~4747 		& 13 $\pm$ 3 		& 27 $\pm$ 3		& 12 $\pm$ 3			& 6 $\pm$ 1				& 12 $\pm$ 2		& 1.1 - 7.3			& 1.1 - 12.0			& 0.052 $\pm$ 0.002\\ 
NE-1			& 1.2 $\pm$ 0.4		& $\leq 1.4$			& [4 $\pm$ 1]			& [$\geq 3$]				& [$\geq 9$]			& 0.57 - 2.3			& 0.57 - 1.8			&0.0014 $\pm$ 0.0007\\
NE-2			& 1.4 $\pm$ 0.4		& $\leq 1.5$			& [4 $\pm$ 2]	 		& [$\geq 3$]				& [$\geq 25$]		& 0.57 - 2.2			& 0.8 - 1.8			& 0.0007 $\pm$ 0.0001\\
SW tail 			&-- 					& --					& --						& --						& --					& 1.3 - 1.6			& 1.3 - 1.6			& 0.003 $\pm$ 0.001\\
NE tail 	 		&-- 					& --					& --						& --						& --					& 1.0 - 2.5			& 1.1 - 3.0 			& 0.014 $\pm$ 0.002\\
Stellar knot - 1  	&-- 					& --					& --						& --						& --					& 0.51 - 1.3			& 0.004 - 0.51		& 0.0023 $\pm$ 0.0004\\
Stellar knot - 2	&-- 					& --					& --						& --						& --					& 0.006 - 1.9 		& 0.51 - 1.8			& 0.0005 $\pm$ 0.0002\\
\hline 
\multicolumn{8}{l}{\textbf{NGC~3166/9:}}\\
NGC~3165		& 2.3 $\pm$ 4 		& 11 $\pm$ 1 		& 33 $\pm$ 6			& 7 $\pm$ 1				& 8.8 $\pm$ 0.6		& 11.3 - 11.5			& 7.5 - 8.0 		& 0.060 $\pm$ 0.001\\
NGC~3166		&-- 					& 590 $\pm$ 10	 	& --						& --						& --					& 4.0				& 5.0  			& 0.06 $\pm$ 0.01\\
C1			 	& 0.4 $\pm$ 0.1 		& $\leq 0.6$ 		& [6 $\pm$ 3]			& [$\geq 4$]				& [$\geq 68$] 	 	& -- 				& --				& $<0.001$\\ 
C2				& 0.6 $\pm$ 0.1 		& $\leq 0.7$ 		& [5 $\pm$ 2]			& [$\geq 4$]				& [$\geq 12$] 	 	& 0.004 - 1.9		& 0.004 - 1.8	& $<0.001$\\ 
AGC~208443 	&1.4 $\pm$ 0.2		& 1.6 $\pm$ 0.3  	& 10 $\pm$ 3			& 9 $\pm$ 2				& 15 $\pm$ 3 		& --					& --				& 0.015 $\pm$ 0.001\\ 
AGC~208444 	& 0.9 $\pm$ 0.2		& 1.1 $\pm$ 0.2 		& 10 $\pm$ 3			& 8 $\pm$ 3				& 21 $\pm$ 5  	 	& -- 				& --				& 0.004 $\pm$ 0.001\\ 
AGC~208457 	& 3.1 $\pm$ 0.4 		& 3.2 $\pm$ 0.4 		& 1.5 $\pm$ 0.4			& 1.4 $\pm$ 0.4			& 15 $\pm$ 4  		& 0.006 - 2.6		& 0.006 - 2.2	& 0.008 $\pm$ 0.001\\ 
C3				& 0.7 $\pm$ 0.2 		& $\leq 2.4$ 		& [4 $\pm$ 2]			& [$\geq 1$]				& [$\geq 9$] 		& --					& --				& $<0.001$\\ 
NGC~3169		& 56 $\pm$ 7	 	& 710 $\pm$ 20		& 54 $\pm$ 8			& 4.3 $\pm$ 0.3 			& 13 $\pm$ 1 		& 7.0 - 11.8			& 10.0 - 11.5  	& 0.56 $\pm$ 0.03\\%
C4				& 0.5 $\pm$ 0.1 		& $\leq 1.4$  		& [19 $\pm$ 8]			& [$\geq 6$]				& [$\geq 16$] 		& 2.1 - 2.4			& 2.1 - 2.3		& $<0.001$\\ 
AGC~208537 	& 0.9 $\pm$ 0.1 		& $\geq 0.9$ 		& 5 $\pm$ 1				& --						& --				 	& --					& --				& --	\\ 
\hline 
\end{tabular}
\end{minipage} 
\end{table*}


\section{Results: Low-mass galaxy populations in NGC~4725/47 and NGC~3166/9}
\label{results}

We have presented high-resolution {H\sc{i}} observations from the GMRT and deep optical imaging from the CFHT MegaCam of two nearby gas-rich groups in our small survey.  Including NGC~871/6/7, which was analyzed in depth in Paper~2, each group in this study contains two interacting spiral galaxies and at least one probable tidal object.  Individually, the low-mass objects in these galaxy groups offer information about the evolutionary history of their respective groups.  This section focusses on the nature of the low-mass galaxies in the NGC~4725/47 and NGC~3166/9 groups. 

\subsection{NGC~4725/47}
\label{results:NGC4725/47}

The GMRT observations reveal that NGC~4725/47 is host to two probable tidal knots (NE-1 and NE-2) and one gas-rich dwarf galaxy (KK~167).  Both knots, which are fully contained within a gaseous tail presumed to have been produced as a result of a recent tidal interaction event, appear to have sufficient mass ($M_{gas}$ $\approx$ 10$^8$ $M_{\odot}$) to evolve into long-lived TDGs according to numerical simulations (e.g.~\citealt{bou2006}).  Nevertheless, NE-1 is located closer to NGC~4747 than NE-2 --- with separation distances of $\sim$4 arcmin (15 kpc) and $\sim$7 arcmin (25 kpc), respectively, from the centre of the host spiral --- causing the former to have a higher likelihood of succumbing to the gravity of its parent galaxy.  We note that NE-1 and NE-2 are unlikely to be self-gravitating and therefore their values of $M_{dyn}$ (and $M_{dyn}/M_{gas}$) may have little meaning.  

Fig.~\ref{fig:NGC4725_cfht}c shows that neither NE-1 nor NE-2 has a clearly bound stellar counterpart, despite the fact that the gas-rich tail is situated along a prominent stellar tail.  This lack of stars could indicate that the {H\sc{i}} tail is fairly young, since significant star formation in a tidal feature typically begins $\sim$20 Myr after being triggered by an interaction event \citep{kav2012}.  Nevertheless, the age ranges estimated for the NE and SW tails do indicate that an interaction likely occurred $\geq 1$ Gyr ago, which is sufficient time for {H\sc{i}} knots to begin forming stars.  Additionally, the FUV component of the NE tail traces the lowest contour of the {H\sc{i}} distribution, as illustrated in right panel of Fig.~\ref{fig:N4747extract}, and indicates active star formation along the edges of the {H\sc{i}} feature, which introduces further complexity.  It is likely that the NE tail is one physical feature and its bifurcated appearance on the sky (in previous optical and UV imaging) is the result of projection effects, similar to the tidal tail presented in \citet{bri+2004}; however, our observations are unable to disentangle the exact geometry of this feature.  

If there is an overlap of the tidal tails along the line of sight, then the sizes and {H\sc{i}} masses for NE-1 and NE-2 could be overestimated, since they could be composed of distinct smaller clouds in the superimposed tails.  These two tidal knots would then be less likely to have bound stellar components and evolve into long-lived TDGs.  We note that while this scenario is a possibility, we find no spectral evidence for distinct clumps along the same line-of-sight (Fig.~\ref{fig:globalprofile}).  Moreover, the {H\sc{i}} mass of the tail measured by ALFALFA well exceeds the mass of NE-1 and NE-2 recovered by the GMRT.  Ignoring possible projection effects, this excess suggests that these knots could continue to accrete significant amounts gas from the tail.  If NE-1 and NE-2 continue to accumulate mass and move a sufficient distance away from NGC~4747 to avoid falling back into the latter, then optimistically, both of these objects could evolve into long-lived TDGs. 

KK~167 is not in the immediate vicinity of another galaxy and appears to be a dark matter dominated dIrr with SFR $= 0.010 \pm 0.001$ $M_{\odot}$ yr$^{-1}$.  Nevertheless, the {H\sc{i}} distribution in KK~167 appears truncated within a faint stellar disk, which is unusual for undisturbed dIrrs (e.g.~\citealt{bro1997}, \citealt{hun2012}).  The optical appearance of KK~167 is similar to other low-mass galaxies such as blue compact dwarfs (e.g.~\citealt{pus2005}) but its SFR also precludes it from this classification (see \citealt{cai2001}).  Although there are clear signs of tidal interactions occurring within the NGC~4725/47 group, it is unclear whether these same interactions are responsible for the stellar properties and truncated {H\sc{i}} distribution of KK~167.  As previously mentioned, the GMRT pointing centred on this source was slightly noisier and required additional flagging of short baselines.  It appears that $\sim 3 \times 10^7$ $M_{\odot}$ of the {H\sc{i}} in KK~167 (that was detected by ALFALFA) is resolved out by the GMRT and is assumed to be diffusely distributed around the galaxy.  Therefore, with additional consideration of the limitations of the \citet{bc2003} stellar age estimates, we conservatively identify KK~167 to be a classical dIrr rather than a tidal feature in the discussion below.


\subsection{NGC~3166/9}
\label{results:NGC3166/9}

NGC~3166/9 contains a TDG candidate, AGC~208457, linked to NGC~3166 by a very faint tidal tail, at least three dIrrs and four {H\sc{i}} knots.  AGC~208457, as shown in Paper~1, has a low dynamical to baryonic mass ratio indicating that little to no dark matter is associated with this feature.  Measurements from the deep optical photometry presented here for this group show that AGC~208457 is bluer, with a 0.28 mag $g$-$r$ colour, and at least a few Gyr younger --- for either a Chabrier or Salpeter IMF and allowing for metallicities in the range of 0.0001 $< Z <$ 0.05 --- than its purported parents, NGC~3166 and NGC~3169 (both with $g$-$r \sim 0.75$ mag).  These results are comparable to those from \citet{kav2012}, who find that tidal knots have a median colour offset of $\sim$0.3 mag in $g$-$r$ from their parents and the average age of interacting spirals is $\sim$7 Gyr.

Gas-phase metallicity measurements, which would ultimately verify tidal origins for AGC~208457 (e.g.~\citealt{duc2007}, \citealt{swe2014}) --- using optical spectroscopy in a similar manner as presented in \citet{duc2014} --- are currently being explored for this feature.  If AGC~208457 has a low metallicity of $Z$ = 0.0001, then it would be 1.6 - 2.6 Gyr in age; whereas, a higher metal content (i.e.~$Z \geq$ 0.004) would indicate that the stars in this object formed within the last Gyr.

As detailed in Paper~1, AGC~208443 and AGC~208444 are likely in-falling dIrr galaxies that are interacting with each other.  The residual reflection halo from a nearby bright star hindered the ability to obtain additional details about the stellar components of these two dwarfs; however, since they are located in the peripheral region of the group, it is unlikely that they have influence on the central group members.  Our {H\sc{i}} and optical observations reveal that AGC~208537 is relatively isolated (i.e.~not located near any tidal features) and has a velocity gradient across its major axis that suggests rotation.  Unfortunately, another foreground star prohibits the detection and measurement of an optical counterpart.  The measured dynamical to gas mass ratio is $\sim$5 for AGC~208537 and since typical dIrrs have $M_{H_I}/M_{stellar} \approx 1$ \citep{bra2015}, any likely associated optical feature would not change the fact that this galaxy is dark matter dominated.

The {H\sc{i}} knots C1 to C4 appear to lie along a gaseous arc that is consistent with a tidal tail stemming from NGC~3169.  The lack of significant stellar counterparts for all four GMRT detections, their location in the tidal tail of NGC~3169 and the young age of the optical region coinciding with two of the {H\sc{i}} knots suggest that these objects are tidal in nature.  Their gas masses ($M_{H_I} \approx 5 \times 10^7$ $M_{\odot}$) are below the threshold required to become long-lived TDGs and would remain as such even if they accrete all of the gas detected by ALFALFA in this region (see Paper 1).  These low masses indicate that C1 to C4 are unlikely to be self-gravitating and will eventually fall back into their parent galaxy within a few 100 Myr (see \citealt{bou2006}).


\section{Discussion: the Frequency and Diversity of Long-lived Tidal Features}
\label{discuss}

With high-resolution {H\sc{i}} observations and deep optical photometry now in-hand for all three galaxy groups in this study, it is now possible to consider the implications of the survey results as a whole.  The consequences of the tidal features detected in the survey are discussed in this section.

Simulations within the standard cosmological framework predict that the properties of TDGs and large tidal knots differ from dIrrs.  A key distinguishing characteristic is the dynamical to baryonic mass ratio \citep{hun+2000}.  Although dynamical mass estimates rely heavily on the assumption of self-gravitation and dynamic equilibrium within a source, TDGs and tidal knots appear to have significantly less total mass --- and therefore little to no dark matter --- compared to their classical counterparts.  Discernment can also arise from age, metallicity and baryonic content; however, measurement of these properties have proven to be quite difficult for low surface brightness objects.  Our main constraint on the origin of the low-mass {H\sc{i}} features detected in this survey relies on the dynamical to baryonic mass ratio for each object.

\begin{figure}
\centering
  \includegraphics[width=84mm]{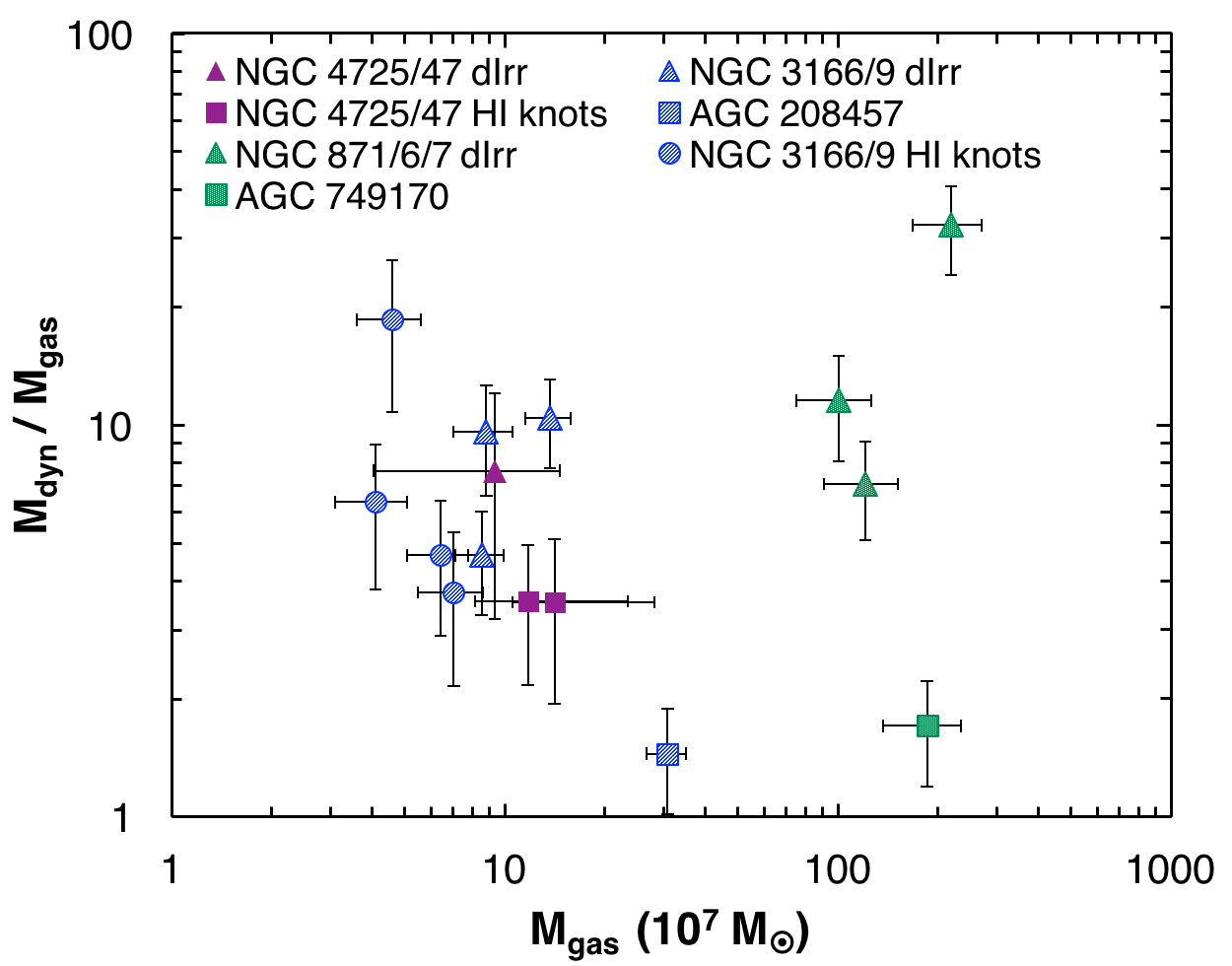}
  \caption[]{Comparison of mass ratios for the low-mass objects detected in this study.  The triangles indicate dIrrs and other first-generation galaxies, squares are potentially long-lived tidal features, and circles are short-lived HI knots.  AGC 208457 and AGC 749170 are found in the lower region of the plot, which is indicative of their tidal nature. 
\label{fig:final_summary}}
\end{figure}

\begin{figure}
\centering
  \includegraphics[width=84mm]{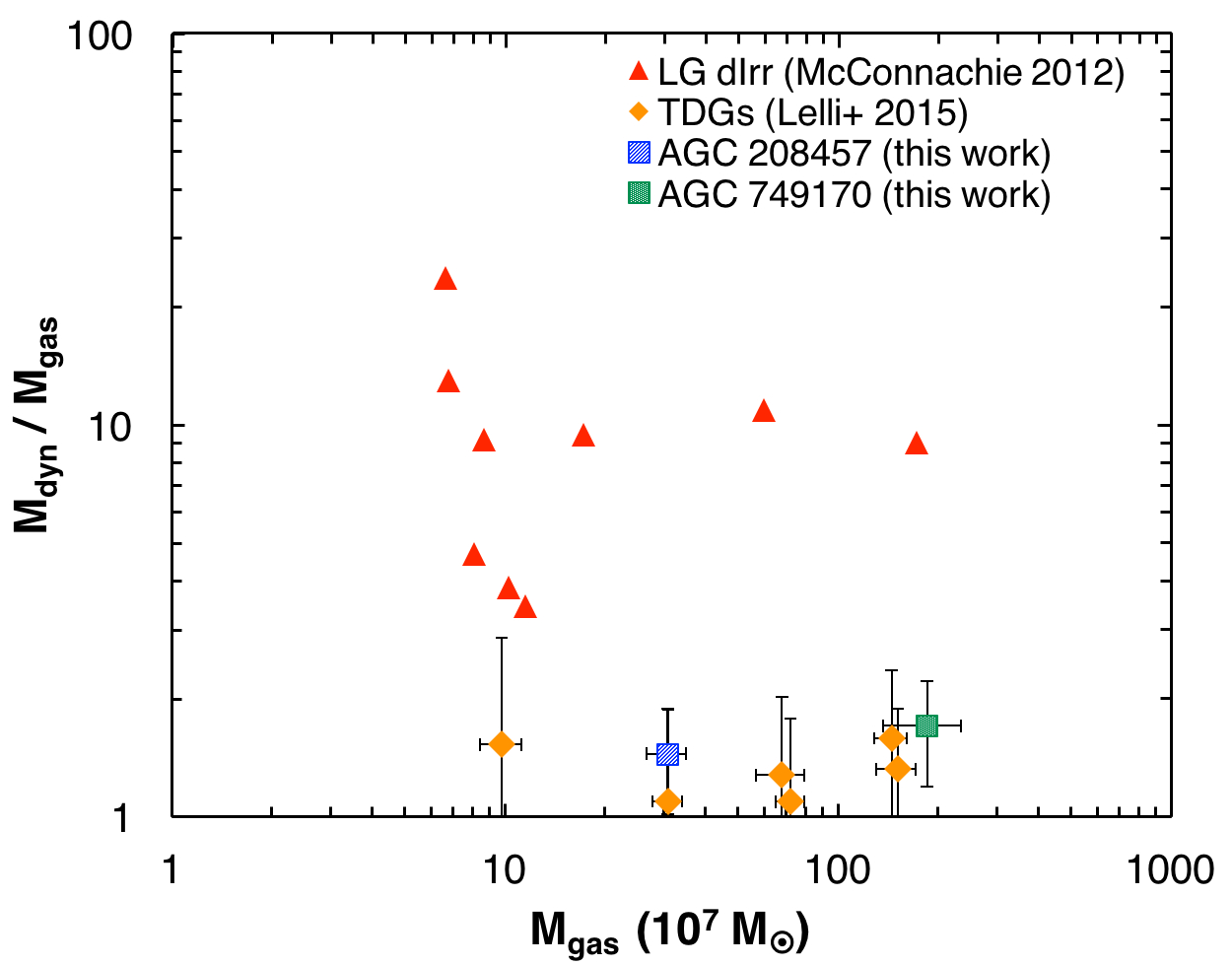}
  \caption[]{Comparison of the large tidal knots (which are likely to be self-gravitating) detected in this study, dIrrs in the LG (as classified by \citealt{mcc2012}) and TDGs reported by \citet{lel2015}.  $M_{gas}$ for each of the LG dIrrs is computed from the data compiled by \citet{mcc2012} (assuming $M_{gas}$ = 1.33$M_{H_I}$) and $M_{dyn}$ is based on the values reported in that same paper, where applicable, or from the total masses reported in \citet{mat1998}.  For the TDGs, $M_{gas}$ is the sum of $M_{H_I}$ and the molecule gas of each object presented in \citet{lel2015}. 
\label{fig:final_compare}}
\end{figure}

Overall, as summarized in Table~\ref{class}, the three groups in our study contain a total of eight spiral galaxies, at least eight dIrrs, four tidal knots (with $M_{H_I}$ $\approx$ 10$^7$ $M_{\odot}$) that are likely short-lived, and four tidal knots containing sufficient gas to survive and evolve into long-lived TDGs.  Fig.~\ref{fig:final_summary} compares $M_{dyn}/M_{gas}$ of all 16 gas-rich low-mass objects detected in the three groups.  AGC 208457 and AGC 749170 appear distinct (i.e.~are found in the lower region of the plot) from the other dwarf galaxies and tidal knots that are currently not self-gravitating.  The mass ratios of these two objects are compared to the properties of classical gas-rich dwarfs compiled by \citet{mcc2012} and TDGs presented in \citet{lel2015} in Fig.~\ref{fig:final_compare}.  There is a clear separation between TDGs and other gas-rich galaxies in this parameter space, with the TDGs (including AGC 208457 and AGC 749170) occupying the lower portion of the plot and dIrrs occupying the upper portion.

\begin{table}
\centering
 \begin{minipage}{84mm}
\caption{Classification of detected gas-rich galaxies and tidal features in all three groups}
\label{class}
\begin{tabular}{l l l} 
\hline
Name 			& Classification 	& Interpretation \\		
\hline
\multicolumn{3}{l}{\textbf{NGC~3166/9:} (\citealt{lee2012}, this work)}\\
NGC~3165 		& spiral			&\\
NGC~3166		& spiral			& interacting with NGC~3169\\
C1 				& tidal knot		& short-lived \\ 
C2 				& tidal knot		& short-lived \\ 
AGC~208443 	& dIrr			& interacting with AGC~208444\\ 
AGC~208444 	& dIrr			& interacting with AGC~208443\\ 
AGC~208457 	& tidal knot		& TDG candidate, long-lived \\ 
C3 				& tidal knot		& short-lived \\ 
NGC~3169 		& spiral			& interacting with NGC~3166\\
C4 				& tidal knot		& short-lived \\ 
AGC~208537 	& dIrr			&\\ 
\hline
\multicolumn{3}{l}{\textbf{NGC~871/6/7:} \citep{lee2014}}\\
AGC~748849	& dIrr			&\\
NGC~871		& spiral			&\\
AGC~121467	& dIrr			&\\
UGC~1761		& dIrr/Irr			&\\ 
AGC~749170 	& tidal knot		& optically dim, long-lived\\
AGC~748853	& dIrr			&\\
NGC~876		& spiral			& interacting with NGC~877\\	
NGC~877		& spiral			& interacting with NGC~876\\
\hline
\multicolumn{3}{l}{\textbf{NGC~4725/47:} (this work)}\\
NGC~4725		& spiral			& interacting with NGC~4747\\ 
KK~167			& dIrr			&\\
NGC~4747 		& spiral			& interacting with NGC~4725\\ 
NE-1			& tidal knot		& potentially long-lived\\
NE-2			& tidal knot		& potentially long-lived\\
\hline 
\end{tabular}
\end{minipage} 
\end{table}

AGC~208457, found in NGC~3166/9, meets most of the criteria for a TDG and is quite similar to another candidate, MDL92, found in NGC~4038/9 (see \citealt{mir1992}).  Both are located at the tip of a tidal tail, contain sufficient {H\sc{i}} mass to become self-gravitating and have an optical counterpart consisting of young stars (see \citealt{hib2001} for a detailed analysis of MDL92).  Nevertheless, neither AGC~208457 nor MDL92 appear to show clear signs of rotation, contrasting the TDGs observed and modelled by \citet{lel2015}.  This discrepancy could be attributed to limited resolution of the observations (considering that two of the six TDGs in \citet{lel2015} have velocity gradients spanning $<$20 km s$^{-1}$) or could indicate that some TDGs are more supported by pressure rather than by rotation.

AGC~749170, in NGC~871/6/7, is a fairly massive {H\sc{i}} knot ($M_{H_I}= 1.4 \pm 0.4 \times 10^{9}$ $M_{\odot}$) with a possible stellar component ($M_{stellar} <10^6$ $M_{\odot}$) comprising fairly young stars (Paper~2).  This extremely optically dim feature appears to be tidal and is located $\sim$90 kpc away from its likely parent galaxies; however, the lack of a readily detectable star forming optical counterpart precludes its classification as a `typical' TDG or tidal knot.  As described in Paper~2, the column density of the {H\sc{i}} in AGC~749170 falls below the star formation threshold (defined by \citealt{sav2004}) and its gas volume density is similar to that of other tidal debris with minimal signs of star formation (see \citealt{may2007}).  

AGC~749170 also has comparable properties to a few other {H\sc{i}}-rich galaxy-like objects (e.g.~\citealt{che1995}, \citealt{eng2010}, \citealt{can2015}), which possibly lie on the extreme end of a class of optically dark/dim tidal features.  Simulations predict that a tidal object this massive and this far from its parent galaxies should be long-lived \citep{bou2006}; however, the stellar counterpart to AGC~749170 is much fainter than any simulated object (e.g.~\citealt{duc2004}).  Optimistically assuming that the longevity predictions for simulated TDGs hold for AGC~749170 as well, then the interaction event in NGC~871/6/7 produced one long-lived TDG.

In NGC~4725/47, the discussion in Section~\ref{results:NGC4725/47} suggests that NE-1 and NE-2 are likely young tidal knots that have yet to enter into their interaction induced star-bursting phases.  Possible projection effects of the tidal tails would indicate over-estimates in the {H\sc{i}} masses for these features.  Nevertheless, these knots could accrete more gas from the tail in which they are embedded and eventually move a sufficient distance from NGC~4747, which would enable them to become self-gravitating TDGs.  If this scenario corresponds to the future of NE-1 and NE-2, then the interaction in NGC~4725/47 will have produced two long-lived TDGs.

To summarize the survey results, our observations of the NGC~3166/9 group reveal one TDG candidate, while optimistic assumptions regarding the future evolution of the {H\sc{i}} features detected in NGC~871/6/7 and NGC~4725/47 suggest that these groups could respectively harbour one and two long-lived TDGs.  The incidence of long-lived TDGs with $M_{baryon} \geq 10^8$ $M_{\odot}$ (and a variety of stellar counterparts) across the three groups studied could therefore be as high as 1.3 TDGs per interacting galaxy pair. 

\subsection{Comparison to Simulations}

Assuming that the two largest spirals in each group produced the tidal objects discussed above, then the incidence and the properties of the latter can readily be compared to TDGs produced in simulations of interacting galaxy pairs.  Numerical simulations by \citet{bou2006} show an average production rate of 6.2 tidal substructures (with $M_{baryon} > 10^8$ $M_{\odot}$) per galaxy interaction event.  The majority ($\sim$65\%) of these features quickly dissipate (i.e.~fall back into their parent galaxies or fade below the detection limit of $M_{baryon} < 10^8$ $M_{\odot}$) within 500 Myr and after 2 Gyr, only 20\% of the tidal knots --- i.e.~$\sim$1.2 tidal knots per interaction event --- remain intact (many of which had original masses $> 10^9$ $M_\odot$).  Alternatively, simulations used by \citet{yan2014} find that a few tens of large tidal knots, with $M_{gas} \geq 10^8$ $M_{\odot}$, are produced during each galaxy interaction event and most of the knots are assumed to be long-lived (surviving over 9 Gyr).

The unbiased survey of low-mass features presented here is sensitive to {H\sc{i}} masses as low as $M_{H_I} \sim 10^7$ $M_{\odot}$ and would therefore detect the tidal objects produced in the various models and simulations discussed above.  We find similar formation and longevity statistics of observable galaxy-like tidal objects to those produced in the simulations by \citet{bou2006}.  Our observations show a total of four potentially long-lived TDGs formed in the three groups studied and that even short-lived tidal knots are not overly abundant in these systems.  

These results do not corroborate the hypothesis that interaction events can produce several tens of viable long-lived TDGs (e.g.~\citealt{yan2014}).  Nevertheless, our sample size is small and was specifically chosen to study the constituents within recently interacting groups, which may not be directly comparable to the longer range simulations.  It is also evident that the simulations of \citet{yan2014} require a much more favourable interaction geometry for producing large tidal knots than those probed in this study in order to reconcile their TDGs formation rates with the incidence of TDGs found here.  

Additionally, the stellar properties of the tidal objects in the three groups do not necessarily resemble the TDGs produced by simulations.  Numerical simulations by \citet{duc+2004} show that TDGs, with masses on the order of $M_{tot} \sim 10^9$ $M_{\odot}$, should contain at least 25\% of their masses in the form of stars, the majority of which were formed in situ.  AGC~208457, in the NGC~3166/9 group, does appear to have a significant stellar component consisting of young stars; however, its stellar mass is still relatively low.  

The critical {H\sc{i}} column density required for star formation to occur is $N_{H_I,crit} = (3-10) \times10^{20}$ atoms cm$^{-2}$ \citep{sch2004}.  NE-1 and NE-2, in the NGC~4725/47 group, are possibly too young to be star-bursting but do have a sufficient average column density --- $N_{H_I,avg} \sim 4 \times 10^{20}$ atoms cm$^{-2}$ for both features as measured from the 30 arcsec resolution GMRT mom$_0$ maps --- to begin forming stars.  It is possible that these two knots will eventually resemble the TDGs produced in simulations.

AGC~749170, in the NGC~871/6/7 group, has an abundant amount of {H\sc{i}} and the absence of a detectable tail indicates that it has had sufficient time to detach from NGC~876 and/or NGC~877 and evolve independently.  This probable tidal object is neither actively star forming nor does it possess an obvious stellar component and therefore it does not resemble the TDGs formed in simulations.  Overall, the TDGs in the groups studied here exhibit a wide range of properties, only some of which can be explained using timing arguments.

The diversity in initial properties and interaction parameters of the parent galaxies in an interaction event can cause significant variations in tidal features that are produced; however, current simulations tend to model pairs of galaxies and do not take into account broader effects of the group environment.  It is possible that the mechanisms that form TDGs are complicated by the gravitational effects of neighbouring galaxies and the intra-group medium, even in the relatively loose groups in this study.  In addition, numerical work by \citet{smi2013} show that due to their lack of dark matter, TDGs should be highly susceptible to ram pressure, which can truncate star formation and strip gas and stellar content.  Determining whether the lack of stars in the tidal objects studied in this survey is the result of initial environmental parameters or post-formation external mechanisms warrants detailed follow-up. 

\section{Conclusions}

We have conducted a detailed multi-wavelength investigation of the gas-rich dwarf galaxy populations of three nearby interacting groups.  The galaxy groups were selected from the blind ALFALFA survey, which allowed detailed follow up on every detection in each group to produce an unbiased census of their dwarf galaxy populations.  High-resolution GMRT observations were used to resolve the {H\sc{i}} belonging to the low-mass group members and measure their dynamical masses (under the strong assumption that each detection is self-gravitating in dynamical equilibrium).  Deep optical imaging from the CFHT provided estimates of the stellar masses and ages of putative optical counterparts to the {H\sc{i}} detections.  The combination of {H\sc{i}} data and optical photometry enabled the distinction between and classification of dIrrs, short-lived tidal knots, and TDG candidates in our unbiased census.  

Taking into account all the features detected in the three groups in this survey, majority of the detectable low-mass objects appear to be either classical dIrr galaxies or short-lived tidal knots.  AGC~208457 --- located in NGC~3166/9 --- has the hallmarks of a standard TDG, albeit without clear signs of rotation.  AGC~749170 --- located in NGC~871/6/7 --- is a gas-rich tidal cloud that has clearly detached from its parent galaxies; however, the lack of a significant stellar component designates AGC~749170 into a possibly rare class of optically dark/dim tidal features that are not currently found in simulations.  In NGC~4725/47, tidal knots NE-1 and NE-2 appear to be in the very early stages of formation and both objects have the potential to evolve into long-lived TDGs.

Overall, four tidal objects with $M_{H_I}\geq 10^8$ $M_{\odot}$ have been identified in the three groups.  These objects will possibly become long-lived tidal galaxies, implying a TDG production rate in agreement to that found by the simulations in \citet{bou2006}, which is considerably lower than the rate implied in the work by \citet{yan2014}.  Nevertheless, the diversity in the properties of tidal features are not reflected in current simulations, even at the high-mass end (i.e.~AGC~749170).  With the advancement of modern telescopes that are beginning to detect even fainter objects, simulations with higher sensitivity and resolution are required to model lower mass features and take into account the broader environmental effects within galaxy groups.

\section*{Acknowledgements}

We thank the staff of the GMRT that made our interferometric observations possible.  Thank-you to the reviewer, P.-A. Duc, for his thoroughly detailed suggestions and comments to improve the clarity of this paper.  We also thank J.A. Irwin for her input on the overall research project.  K.S. acknowledges funding from the National Sciences and Engineering Research Council of Canada.   The ALFALFA team at Cornell is supported by U.S. NSF grants  AST-0607007 and AST-1107390 to R.G. and M.P.H. and by grants from the Brinson Foundation. The GMRT is run by the National Centre for Radio Astrophysics of the Tata Institute of Fundamental Research.  This research used optical observations obtained with MegaPrime/MegaCam, a joint project of CFHT and CEA/DAPNIA, at the Canada-France-Hawaii Telescope (CFHT) which is operated by the National Research Council (NRC) of Canada, the Institute National des Science de l'Univers of the Centre National de la Recherche Scientifique of France, and the University of Hawaii.







\bsp	
\label{lastpage}
\end{document}